\documentclass[a4paper,11pt]{article}
\pdfoutput=1 

\usepackage{jheppub} 

\usepackage[T1]{fontenc} 
\usepackage{graphicx,amssymb,amstext,amsmath}
\usepackage{subcaption}
\usepackage{caption}
\usepackage{appendix}
\newcommand{\be}{\begin{equation}}
\newcommand{\ee}{\end{equation}}
\newcommand{\p}{\partial}

\title{\boldmath Comments on the stability of the KPV state}

\author{Nam Nguyen}
\affiliation{Department of Mathematical Sciences and Centre for Particle Theory,\\Durham University, Durham DH1 3LE, United Kingdom}

\emailAdd{nam.h.nguyen@durham.ac.uk}

\abstract{Using the blackfold approach, we study the classical stability of the KPV (Kachru-Pearson-Verlinde) state of anti-D3 branes at the tip of the Klebanov-Strassler throat. With regards to generic long-wavelength deformations considered, we found no instabilities. We comment on the relation of our results to existing results on the stability of the KPV state.}

\begin{document}
\maketitle
\flushbottom

\section{Introduction}
Understanding the dynamics of antibranes in fluxed background, particularly anti-D3 branes in Klebanov-Strassler background, has been of revived interest in recent years. This is due, in part, to the debate over the validity of the KKLT (Kachru, Kallosh, Linde, Trivedi) construction of de-Sitter vacua \cite{Kachru2003DeTheory}, in which metastable state of anti-D3 branes remains a controversial prerequisite \cite{Danielsson:2018ztv}. 

\paragraph{A brief review}
The KPV (Kachru-Pearson-Verlinde) state \cite{Kachru2002Brane/fluxTheory} is a proposed configuration of anti-D3 branes at the tip of the Klebanov-Strassler \cite{Klebanov2000SupergravitySingularities} background. Originally in \cite{Kachru2002Brane/fluxTheory}, it was argued that anti-D3 branes can polarise into a spherical NS5 brane, and that, in probe approximation, in the regime of $p/M$\footnote{$p$ denotes the number of the anti-D3 branes and $M$ the strength of the Klebanov-Strassler background flux.} between 0 and $p_{crit}$ with $p_{crit} \approx 0.080488$, the polarised anti-D3-NS5\footnote{NS5 branes with dissolved anti-D3 brane charge.} brane balances its own ``weight'' with ``electromagnetic'' forces from the fluxes to form a metastable configuration.  

Because of the singularities found when considering backreaction of anti-D3 branes to the throat \cite{Bena2010OnKlebanov-Strassler}, concerns about the existence of the state were raised. It’s important to note that the KPV state is actually an anti-D3-NS5 state formed by the polarisation of anti-D3 branes under non-trivial fluxes as opposed to a state of localised anti-D3 branes. Nevertheless, the singularities still mean bad news for KPV especially when \cite{Bena:2014jaa} pointed out that the KPV state is outside of the regime of validity of probe analysis.

The first evidence in favour of the existence of the KPV state came in the form of \cite{Michel:2014lva}, where, through consideration of a single anti-D3 brane, it was argued that there exists a \textit{possibility} that the previously found singularities can be avoided. However, this possibility was later explored in \cite{Bena:2016fqp} where it was shown that, at least in certain regime of parameters, the possibility cannot be realised. Subsequently, \cite{Cohen-Maldonado:2015ssa} observed that singularities are not expected to appear once we consider \textit{extremal} anti-D3-NS5 branes. Treating backreaction perturbatively through the blackfold approach, \cite{Armas:2018rsy} showed evidence of the existence of the KPV configuration  \textit{exactly} where no-go theorems are evaded. More precisely, it was found that the polarised anti-D3-NS5 branes could form a metastable state at the tip of the throat, and such solution would disappear as soon as we heat up the polarised state \textit{sufficiently} that it geometrically resembles localised black anti-D3 branes.

\paragraph{Our focus}
It is important to note that the claim regarding metastability of the relevant anti-D3-NS5 state in \cite{Kachru2002Brane/fluxTheory} and subsequently in \cite{Armas:2018rsy} is only with respect to \textit{some} modes of deformations and not a general statement of stability. For example, in \cite{Armas:2018rsy}, only spherically homogeneous transformations were considered. This means spherically non-homogeneous deformations of the KPV state were ignored. For the purpose of cosmological model building through uplifting, we need not only that the configuration exists but also that it is long lived. However, there is evidence suggesting that this might not be the case, at least for certain regimes of parameters.

In \cite{Bena:2014jaa}, from the perspective of localised anti-D3 branes, it was argued that there exists a direction along which the branes feel repulsive forces among themselves and destabilise away from the KPV state. This suggests that, in appropriate regime of parameters, the KPV configuration suffers from fragmentation instability. 

From the complementary perspective of anti-D3-NS5 branes, we study the stability properties of the KPV state using the blackfold approach. Before continuing, let us stress what our analysis does \textit{not} do. As blackfold is based on the idea of matched asymptotic expansion, one need to specify a seed metric as the description of the solution in the near zone. By choosing the stacked anti-D3-NS5 branes solution as the near zone seed, we have effectively ignored all brane splitting and fragmentation deformations. Moreover, as noted in \cite{Armas:2018rsy}, the analysis is reliable when $p/M$ is not too close to zero, at which point the NS5 brane shrinks and the localised anti-D3 perspective becomes the better description. Since the analysis in \cite{Bena:2014jaa} is done from the localised anti-D3 branes perspective and the discovered instabilities are brane splitting instabilities, the blackfold results presented here should be thought of as complimentary and not contradictory to that of \cite{Bena:2014jaa}. Another important caveat is that, as blackfold theory is an effective theory of long-wavelength interactions, claim of stability is made only with respect to long-wavelength perturbations. For more discussions of the blackfold approach as an effective theory, we refer readers to \cite{Niarchos:2015moa}.

Let us note that a preliminary study of the stability of the KPV state was done in \cite{Bena:2015kia} where it was argued that the spherical NS5 shell is unstable under perturbations. While keeping in mind that the regime of validity of the analysis done in \cite{Bena:2015kia} and of ours are different, as we shall see shortly, our results do not support the picture proposed in \cite{Bena:2015kia}.  

\paragraph{Our results}
Introducing generic long-wavelength worldvolume dependent deformations to the blackfold description of the KPV state, we observe that the blackfold equations (constraints on long-wavelength deformations) prohibit the existence of tachyonic modes. It's interesting to mention also that counter-intuitively, the KPV state, a polarised state of anti-D3 branes, can feel an electromagnetic repulsion away from the tip of the Klebanov-Strassler throat. Nevertheless, this electromagnetic repulsion is ``out-weighted'' by the gravitational pull so the KPV state is still stabilised radially by a net force downward.\footnote{In the previous version, which does not include warping effects of the Klebanov-Strassler throat, we observe a window of instability near $p_{crit}$. In presence of these effects, the window of instability no longer exists.}

\paragraph{Outlook}
Although not discussed in this paper, generalisation of the stability analysis to account for non-extremal branes can be achieved with the same method. If the fragmentation instability is observed for extremal KPV states (in a full analysis of the system, perhaps beyond the method of this paper), then it would be interesting to study thermal effects to see if it is resolved. This possibility is one we would like to pursue in later works.

\paragraph{Outline of paper}

The plan of the paper is as follows. A short derivation of the KPV state from blackfold analysis is reviewed in section \ref{sec2}. The blackfold stability analysis of the KPV state is presented in section \ref{sec3}. A discussion of the Klebanov-Strassler background near the apex is provided in appendix \ref{KS}. Details on the construction of the equivalent currents used in the KPV state derivation is collected in appendix \ref{secA}. Lastly, the derivation of blackfold perturbation equations used in the stability analysis of the KPV state is summarised in appendix \ref{secB}.

\section{KPV state from blackfold}\label{sec2}
\paragraph{Overview} Blackfold theory \cite{Emparan:2009cs,Emparan:2009at,Armas:2016mes} is a long wavelength effective theory of gravity, conceptually based on the technique of matched asymptotic expansions. As a thorough discussion of blackfold and its application to antibranes metastable state has already been given in \cite{Armas:2018rsy} and \cite{M2M5brane}, we shall not repeat it here. Nevertheless, let us briefly present the fundamental of the blackfold argument for the existence of metastable antibranes.

The blackfold equations are the constraint equations of the backreacted metric and gauge fields that match the anti-D3-NS5 branes in the near zone and asymptote the Klebanov-Strassler background in the far zone to first order in a derivative expansion. Analogous to the fluid equations of the Fluid/Gravity correspondence \cite{Bhattacharyya:2008jc}, because of the interplay between derivative expansion and constraint equations, the blackfold equations will determine the zeroth order terms of the derivative expansion. By explicitly solving the blackfold equations, we have proven the necessary conditions for the existence of the KPV state. 

In general, one might be worried that solving the constraint equations alone does not automatically guarantee a full solution. However, in all examples of matched asymptotic expansions that have been worked out in details (most notably \cite{Camps:2012hw}), the constraint equations not only provide the necessary conditions but also the \textit{sufficient} conditions for a regular solution to first order in derivative expansion. It is therefore natural to  \textit{conjecture} that there is a one to one correspondence between a solution of the blackfold equations and a regular solution of the gravitational equations. This conjecture is almost analogous to the statement in Fluid/Gravity that there is a one to one map between a solution of the fluid equations and a regular solution of the gravitational equations.

The purpose of this section is to provide a brief derivation of the KPV state from the blackfold approach. Various aspects of anti-D3-NS5 blackfold, including the recovery of the KPV state, have already been discussed in \cite{Armas:2018rsy}. Nevertheless, we find it useful to revisit the starting point of our stability analysis. We will also take this opportunity to state our conventions, provide some relevant details and explanations, and fix some typos in the literature.

\paragraph{Conventions}
\begin{enumerate}{}
\item The signature is mostly plus $(- + + + ...)$.

\item The Hodge star operator of a $p$-form on an $n$-dimensional manifold is defined as 
\be
(* A)_{\mu_1 ... \mu_{n-p}} = \frac{1}{p!} \epsilon_{\nu_1 ... \nu_p \mu_1 .... \mu_{n-p}} A^{\nu_1 ... \nu_p}
\ee
with $\epsilon_{\nu_1 ... \nu_p \mu_1 .... \mu_{n-p}}$ the Levi-Civita tensor. 

\item Gauge invariant field strengths are defined as
\be
\tilde{F}_{q+2} = F_{q+2} - H_3 \wedge C_{q-1} 
\ee

with the exception of the self-dual $\tilde{F}_5$ which is defined as
\be
\tilde{F}_5 = F_5 + B_2 \wedge F_3
\ee
where $F_{q+2} \equiv d C_{q+1}$.

\item Electric currents appear with a $-$ sign in the forced Maxwell equations: 
\be
d \star F_{p+2} = - 16 \pi G \ J_{p+1}
\ee

\item Magnetic currents appear with a $+$ sign in the forced Maxwell equations: 
\be
d F_{p + 2} = 16 \pi G  \ j_{n - q - 3} 
\ee

\end{enumerate}

\paragraph{Klebanov-Strassler throat}
We refer readers to Appendix \ref{KS} for a complete description of the Klebanov-Strassler background near the apex. For the purpose of deriving the KPV state, we shall only present here the metric and the flux components that contribute to the derivation. As the dilaton of the Klebanov-Strassler solution is a constant, we shall set $g_s = 1$ for our convenience. As discussed in the appendices, the Klebanov-Strassler metric near the apex is given by
\begin{multline}
\label{200}
g_{\mu \nu} d x^\mu d x^\nu = M b_0^2  \Big ( - dt^2 + (dx^1)^2 + (dx^2)^2 + (dx^3)^2  + dr^2 \\
+ d \psi ^2 + \sin^2 \psi \left( d \omega^2 + \sin^2 \omega d \varphi^2 \right)  + r^2 (d \tilde{\omega}^2 +  \sin^2 \tilde{\omega} d \tilde{\varphi}^2) \Big) + ... 
\end{multline}
and relevant fluxes are given by
\begin{align}
F_3 &= 2 M \sin^2 \psi \sin \omega \ d \psi \wedge d \omega \wedge d \varphi + ... \\
H_7 &= - 2 M^3 b_0^4 \sin^2 \psi \sin \omega \ d t \wedge dx^1 \wedge dx^2 \wedge dx^3 \wedge d \psi \wedge d\omega \wedge d \varphi + ...
\end{align}
where $b_0^2 \approx 0.93266$ and the dots refer to components of the metric/flux that do not contribute in our derivation.\footnote{Some terms in the dots are important to our stability analysis and shall be discussed appropriately later on.  
}

\paragraph{Anti-D3-NS5 branes} 
As demonstrated in the literature, the blackfold equations can be obtained as the conservation equations of equivalent sources induced by the branes onto the background in the far zone. Therefore, to obtain the anti-D3-NS5 blackfold equations, one could go to the far zone and ask what equivalent sources can mimic the effects of these branes. In the interest of time and space, let us relegate the details of this process to Appendix \ref{secA} and simply present the results here. For the extremal anti-D3-NS5 branes, we have the equivalent energy-stress tensor
\be
\label{33}
T^{ab} = C \left( - r_h^2 \sin^2 \theta (\gamma^{ab} - v^a v^b - w^a w^b) - r_h^2 \cos^2 \theta \gamma^{ab}  \right)
\ee
and the equivalent currents
\begin{align}
J_2 &= C r_h^2 \sin \theta \cos \theta \ v \wedge w \\
J_4 &= C r_h^2 \sin \theta  * ( v \wedge w) \\
j_6 &= - C r_h^2 \cos \theta  * 1
\end{align}
where $C = \displaystyle \frac{\Omega_3}{8 \pi G}$, and $*$ is the worldvolume Hodge dual operator. 
\vspace{2pt}

\paragraph{Blackfold equations}
In the blackfold set-up of extremal anti-D3-NS5 branes in Klebanov-Strassler background, the variables of the system are
\be
r, \tilde{\omega}, \tilde{\varphi}, \psi, r_h,\tan \theta, v^a, w^a  
\ee    
The variables $r, \tilde{\omega}, \tilde{\varphi}, \psi$ are the embedding degrees of freedom of the anti-D3-NS5 branes to the background. The variables $r_h,\tan \theta, v^a, w^a$ are the characteristic degrees of freedom describing the horizon length, the charge distribution, and the flow of the dissolved charge.\footnote{The intuition for the variables $r_h$ and $\tan \theta$ can be obtained from considering the D3-NS5 supergravity solution in \ref{D3-NS5Sugra}, in which $r_h$ is the extremal horizon radius and $\tan \theta$ is the ratio of D3 brane charge density over NS5 brane charge density $\mathcal{Q}_3/\mathcal{Q}_5$.}

As noted in the introduction, the blackfold equations will describe the \textit{zeroth order} terms in the derivative expansion of the metric and gauge fields that asymptote the stacked anti-D3-NS5 branes in the near zone and the Klebanov-Strassler background in the far zone. These zeroth order terms are obtained from promoting the variables to slowly varying functions of the worldvolume coordinates $\sigma$. For the purpose of describing the KPV configuration, as we are only interested in static and spatially homogeneous configurations of anti-D3-NS5 branes at the tip of Klebanov-Strassler throat, we can already fix variables $r, \tilde{\omega}, \tilde{\varphi}, v^a, w^a$ (see equations (\ref{80})-(\ref{81}) for detailed expressions) and set the remaining variables $r_h, \psi, \tan \theta$ to be constant with respect to the worldvolume coordinates. In our conventions, the blackfold equations are given by\footnote{For the definitions of the geometric quantities used here, one can see Appendix \ref{secB}.}
\begin{enumerate}
\item The energy-momentum conservation equations
\begin{align}
\label{3000}
\nabla_a T^{a b} &= \p^b X_\mu \, \mathcal{F}^\mu \\
\label{3001}
T^{ab} K_{ab}^{\,\,\, \,\,\, (i)}  &= \mathcal{F}^\mu \, n^{(i)}_\mu
\end{align}
where $n_\mu^{(i)}$ denotes the normal vectors of the anti-D3-NS5 blackfold, $K_{ab}^{\,\,\, \,\,\, (i)} = K_{ab}^{\ \ \, \rho} n_\rho^{(i)}$,  and the force term $\mathcal{F}^\mu$ is given by
\begin{multline}
\label{90}
\mathcal{F}^\mu = - \frac{1}{6!}  H_{7}^{\mu a_1 ... a_6} j_{6 a_1 ... a_6} + \frac{1}{2!} \tilde{F}_3^{\mu a_1 a_2} J_{2 a_1 a_2} + \frac{3}{4!} H_3^{\mu a_1 a_2} C_2^{a_3 a_4}J_{4 a_1 ... a_4} \\
+ \frac{1}{4!} \tilde{F}_5^{\mu a_1 ... a_4} J_{4 a_1 ... a_4}  
\end{multline}
For the purpose of describing the KPV state, the terms with $H_3$ and $\tilde{F}_5$ are not relevant because they vanish at the tip of the throat. Nevertheless, as they will play a role when we consider perturbations away from the tip, we present them explicitly here.

\item The current conservation equations 
\begin{align}
\label{34}
d * j_6 &= 0 \\
d * J_4 + * j_6 \wedge F_3  &= 0 \\
\label{35}
d * J_2 + H_3 \wedge * J_4 &= 0 
\end{align}
where $F_3, H_3$ are the projected background fluxes and $*$ is the 6-dimensional Hodge dual of the worldvolume directions.
\end{enumerate}

From the current conservation equations, we can define the conserved Page charges $\mathbb{Q}_3$ and $\mathbb{Q}_5$ that keep track of the number of anti-D3 branes and NS5 branes:
\begin{align}
\mathbb{Q}_5 &= * j_6 = C r_h^2 \cos \theta \\
\label{71}
\mathbb{Q}_3 &= \int_{S^2} * \left( J_4 + *(* j_6 \wedge C_2 ) \right) \\
&= -  4  \pi \Big( C r_h^2 \sin \theta M b^2_0 \sin^2 \psi + C r_h^2 \cos \theta M (\psi - \frac{1}{2} \sin 2 \psi  ) \Big)  
\end{align}
where we have used $C_2 = M (\psi - \frac{1}{2} \sin 2 \psi) \sin \omega d \omega \wedge d \varphi$. It follows immediately that we can write $\tan \theta$ as
\be
\label{1}
\tan \theta = \frac{1}{b_0^2 \sin^2 \psi} \left(  \frac{\pi p}{M} - \left(\psi - \frac{1}{2}\sin 2 \psi \right) \right)
\ee
where we have made the identification
\be
\label{1010}
\frac{- \mathbb{Q}_3}{4 \pi \mathbb{Q}_5} = \pi p
\ee
From the energy-momentum tensor conservation equations, after some algebraic acrobatics, we can write all variables in term of $\psi$ and obtain the equation 
\be
\label{2}
\cot \psi - \frac{1}{b_0^2} \sqrt{1 + \tan^2 \theta} - \frac{1}{b_0^2} \tan \theta = 0
\ee
Integrating equation (\ref{2}) gives us the KPV potential originally obtained from the DBI approach in \cite{Kachru2002Brane/fluxTheory}.   

\paragraph{The KPV state} 
We can numerically determine that equation (\ref{2}) has a metastable solution for $0 < p/M < p_{crit}$ where $p_{crit} \approx 0.080488$. These metastable solutions are the KPV states. For our convenience later, let us note down some explicit information of the configuration. With respect to our variables, the KPV states are specified by
\begin{align}
\label{80}
r &= 0, & \psi &= \psi_0,& \tan \theta &= \frac{1}{b_0^2 \sin^2 \psi_0} \left( \frac{\pi p}{M} - \psi_0 + \frac{1}{2}\sin (2 \psi_0)\right) 
\end{align}

\begin{align}
\label{81}
r_h &= \sqrt{\frac{\mathbb{Q}_5}{C \cos \theta }}, &v^a \p_a &=  \frac{1}{\sqrt{ M} b_0 \sin \psi_0 } \partial_\omega ,& w^a \p_a &= \frac{1}{\sqrt{M} b_0 \sin \psi_0 \sin \omega} \partial_\varphi 
\end{align}
where $\psi_0$ is the metastable solution of
\be
\cot \psi - \frac{1}{b_0^2} \sqrt{1 + \tan^2 \theta} - \frac{1}{b_0^2} \tan \theta = 0
\ee
We note also the induced metric on the worldvolume of the anti-D3-NS5 branes
\begin{equation}
\gamma_{a b} d\sigma^a d\sigma^b = M b_0^2 \left( - dt^2 + (dx^1)^2 + (dx^2)^2 + (dx^3)^2  + \sin^2 \psi_0 \left( d \omega^2 + \sin^2 \omega d \varphi \right)  \right ) \, ,
\end{equation}
the non-zero components of the worldvolume Christoffel symbol $\Theta^a_{bc}$
\begin{align}
\label{01}
\Theta^{\varphi}_{\omega \varphi} &= \Theta^{\varphi}_{\varphi \omega} = \cot \omega & \Theta^{\omega}_{\varphi \varphi} &= - \cos \omega \sin \omega  \, ,
\end{align}
the relevant components of the  background Christoffel symbol $\Gamma^{\mu}_{\alpha \beta}$
\begin{align}
&\Gamma^{\psi}_{\omega \omega} = - \cos \psi_0 \sin \psi_0 & &\Gamma^{\psi}_{\varphi \varphi} = - \cos \psi_0 \sin \psi_0 \sin^2 \omega
\end{align}
and the non-zero component of the extrinsic curvature $K_{ab}^{\ \ \, \rho}$
\begin{align}
\label{03}
K_{\omega \omega}^{\ \ \ \psi} &= - \cos \psi_0 \sin \psi_0 & K_{\varphi \varphi}^{\ \ \ \psi} &= - \cos \psi_0 \sin \psi_0 \sin^2 \omega  \, .
\end{align}

\paragraph{Regime of validity} 
Starting from a seed solution, the blackfold approach aims to add long wavelength deformations to the seed in such a way that yields a perturbative solution with the background asymptotics. This process is only possible if the scale of the seed is much smaller than the scale of the background. In the case of anti-D3-NS5 seed and KS background, this translates to the condition
\be
\label{1020}
r_h \ll \sqrt{M} \sin \psi_0
\ee
It's easy to see that, as long as $\psi_0$ is not too close to $0$, this condition can always be satisfied with a large enough $M$. From the description of the KPV state above, we see that $\psi_0$ is finite for all KPV configurations except for when one push $p/M$ parametrically close to zero, at which point $\psi_0$ also goes very close zero. Let us note further that, because of our definition of $p$ in (\ref{1010}), the parameter $p/M$ remains finite even when $M$ is very large.

This concludes the review of the KPV state from the blackfold approach. We refer readers to \cite{Armas:2018rsy} for more information on the derivation as well as discussions on other aspects of the KPV state.

\section{Stability of KPV state}\label{sec3}
The goal of this section is to analyse generic deformations of the KPV configuration. Starting with the blackfold description of the configuration, we introduce generic perturbations by varying slightly all its variables. As the blackfold equations provide the necessary conditions for the perturbed configuration to be a legitimate solution, we shall use the blackfold equations to constrain allowed perturbations. We shall see that, with respect to deformations amendable to the blackfold description, unstable modes are not allowed. 
\subsection{Perturbation parameters}
\label{PerPara}
To introduce perturbations to our system, we vary slightly the variables of the configuration around their KPV values. Explicitly, we have
\begin{align}
r &= 0 + \delta r,& \psi &= \psi_0 + \delta \psi, & r_h &= \sqrt{\frac{\mathbb{Q}_5}{C \cos \theta(\psi_0)}} + \delta r_h,
\end{align}
\begin{align}
\tan \theta = \frac{1}{b_0^2 \sin^2 \psi_0} \left( \frac{\pi p}{M} - \psi_0 + \frac{1}{2}\sin (2 \psi_0)\right) + \delta \tan \theta 
\end{align}
\begin{align}
v^a \p_a &=  \frac{1}{\sqrt{ M} b_0 \sin \psi_0 } \partial_\omega + \delta v^a \p_a ,\\
w^a \p_a &= \frac{1}{\sqrt{ M} b_0 \sin \psi_0 \sin \omega} \partial_\varphi + \delta w^a \p_a
\end{align}
where all variations are functions of the worldvolume coordinates, e.g. $\delta r_h (\sigma)$. To simplify our syntax, from here on we shall denote the variable values at the KPV configuration by the variables themselves, e.g. $\psi_0$ will be denoted as $\psi$, the value of $\tan \theta$ at KPV is denoted as $\tan \theta$, etc. 

Let us make use of symmetries and constraints to minimise the number of parameters we work with while still preserve all the relevant information for the stability analysis. Firstly, because of Lorentz symmetry of the blackfold equations and the original KPV configuration, without loss of generality, we can consider variations involving the worldvolume coordinate $t$ only instead of the full Minkowskian coordinates $t,x^1,x^2,x^3$. Secondly, using the unitary constraints on $v$ and $w$, i.e. $v^a v_a = w^a w_a = 1$, we can show that
\begin{align}
\label{5}
\delta v^\omega &= - \frac{\cos \psi }{\sqrt{M} b_0  \sin^2 \psi} \delta \psi  \\
\label{6}
\delta w^\varphi &= - \frac{\cos \psi }{\sqrt{M} b_0  \sin^2 \psi \sin \omega} \delta \psi 
\end{align}
Thirdly, as we use $v$ and $w$ together as normal vectors to specify the anti-D3 charge flow inside the NS5 branes, it's obvious that we have a rotational gauge symmetry here. Making use of this gauge symmetry along with the orthogonality constraint, i.e. $v^a w_a = 0$, we can set 
\be
\delta v^\varphi = \delta w^\omega = 0
\ee
With the simplifications noted above, our relevant variation parameters are
\begin{gather}
\delta r (t, \omega, \varphi), \delta \psi (t, \omega, \varphi), \delta r_h (t, \omega, \varphi), \delta \tan \theta (t, \omega, \varphi),\\ 
\delta v^t (t, \omega, \varphi), \delta v^\omega (t, \omega, \varphi), \delta w^t (t, \omega, \varphi), \delta w^\varphi (t, \omega, \varphi)
\end{gather}
where $\delta v^\omega, \delta w^\varphi$ can be written in term of $\delta \psi$ as expressed in (\ref{5})-(\ref{6}).

\subsection{Blackfold perturbation equations}
In this subsection, we present the blackfold equations for perturbations around the KPV state. We relegate the exciting details on the derivation of these equations to appendix \ref{secB}. 
\subsubsection{Conservative Currents \& Charges}
As shown in (\ref{A}), the $j_6$ conservation equation implies 
\be
\p_a \, \delta \mathbb{Q}_5 = 0
\ee
where $\mathbb{Q}_5 = C r_h^2 \cos \theta$. This means $\delta \mathbb{Q}_5$ is a constant of motion. Recall that $\mathbb{Q}_5$ keeps track of the number of NS5 branes. As we are interested in the dynamical stability of the KPV configuration, we impose the condition that $ \delta \mathbb{Q}_5$ vanishes. Note that the imposition $ \delta \mathbb{Q}_5 = 0$ automatically fixes $\delta r_h$ in term of $\delta \tan \theta$
\be
\label{002}
\delta r_h = \frac{1}{2} r_h \cos \theta \sin \theta \delta \tan \theta \, .
\ee
As shown in (\ref{B}), the $J_4$ conservation equation implies 
\begin{multline}
\label{007}
- \mathbb{Q}_5 M b_0^2 \sin^2 \psi \sin \omega
\Bigg(  \p_t \delta \tan \theta + 2 \tan \theta \cot \psi \p_t \delta \psi + \frac{2}{b_0^2}  \p_t \delta \psi \Bigg) \\
=  \mathbb{Q}_5  M^{3/2} b^3_0 \tan \theta  \sin \psi \Big( \p_\varphi \delta w^t + \p_\omega  \left( \sin \omega \delta v^t \right) \Big)
\end{multline}
Integrating over $\omega$ and $\varphi$ and enforcing the periodicity conditions
\be
\delta w^t|_{\varphi =0} = \delta w^t|_{\varphi = 2 \pi}
\ee
we obtain\footnote{Let us note that the equation keeps constant the $\mathbb{Q}_3$ Page charge while put no restrictions on the $\mathcal{Q}_3$ brane charge, which is free to vary.}
\be
\p_a \delta \mathbb{Q}_3 = 0
\ee
where
\be
\label{100}
\delta \mathbb{Q}_3 = \int_{S^2} \delta \left( * \tilde{J}_4 \right) = -  \mathbb{Q}_5 M b_0^2 \sin^2 \psi \int d \omega d \varphi \sin \omega \Bigg(  \delta \tan \theta + 2 \left( \tan \theta \cot \psi + \frac{1}{b_0^2} \right) \delta \psi \Bigg)
\ee
This means $\delta \mathbb{Q}_3$ is a constant of motion. In a similar fashion to how the $ \mathbb{Q}_5$ charge keeps track of the number of NS5 branes, the $ \mathbb{Q}_3$ charge keeps track of the number of anti-D3 branes. As we are interested in the dynamical stability of the KPV configuration, we shall impose that $\delta  \mathbb{Q}_3 = 0$. However, note that unlike the $ \mathbb{Q}_5$, the imposition $\delta  \mathbb{Q}_3 = 0$ doesn't automatically guarantee the satisfaction of the current perturbation equation. \\
Finally, as shown in (\ref{C}), the $J_2$ conservation equation implies 
\begin{align}
\label{003}
\cot \theta \cos^2 \theta \p_\omega \delta \tan \theta + \sqrt{M} b_0 \sin \psi \p_t \delta v^t  = 0\\
\cot \theta \cos^2 \theta  \p_\varphi \delta \tan \theta +  \sqrt{M} b_0 \sin \psi \sin \omega \p_t \delta w^t = 0 \\
\label{004}
\p_\varphi \delta v^t - \p_\omega (\sin \omega  \delta w^t )= 0
\end{align}

\subsubsection{Energy-momentum conservation equations}
Recall from (\ref{3000})-(\ref{3001}), the intrinsic and extrinsic blackfold equations 
\begin{align}
\nabla_a T^{a b} &= \p^b X_\mu \, \mathcal{F}^\mu \\
T^{ab} K_{ab}^{\,\,\, \,\,\, (i)}  &= \mathcal{F}^\mu \, n^{(i)}_\mu
\end{align}
Focusing on perturbations around the KPV state, as shown in (\ref{D}), the intrinsic equation implies for $b = t,\omega, \varphi$ respectively
\begin{enumerate}
\item The $t$ intrinsic perturbation equation  
\begin{multline}
\label{005}
\p_t \delta \tan \theta +  \frac{\sqrt{M} b_0}{\sin \psi} \tan \theta \left( \p_\omega \delta v^t + \frac{1}{\sin \omega} \p_\varphi \delta w^t + \cot \omega \delta v^t \right) 
 \\
 + 2 \left( \cot \psi \tan \theta + \frac{1}{b_0^2} \right) \p_t \delta \psi = 0
\end{multline} 

\item The $\omega$ intrinsic perturbation equation 
\be
\label{9}
\sqrt{M} b_0 \sin \psi  \tan^2 \theta  \p_t \delta v^t +   \sin \theta \cos \theta  \p_\omega \delta \tan \theta    = 0
\ee

\item The $\varphi$ intrinsic perturbation equation 
\be
\label{10}
\sqrt{M} b_0 \sin \psi \sin \omega \tan^2 \theta \p_t \delta w^t  + \sin \theta \cos \theta \p_\varphi \delta \tan \theta  = 0
\ee
\end{enumerate}
Similarly, as shown in (\ref{E}), the extrinsic blackfold equation implies
\begin{enumerate}
\item The $\psi$ extrinsic perturbation equation
\be
\label{008}
(\p_t)^2 \delta \psi -  \frac{ \cos^2 \theta}{ \sin^2 \psi}   \nabla^2 \delta \psi  = \frac{2  \cos^2 \theta}{ \sin^2 \psi}  \delta \psi + \frac{2}{b_0^2}  \cos^2 \theta \left( 1 + \sin \theta \right) \delta \tan \theta
\ee

\item The $r$ extrinsic perturbation equation 
\begin{multline}
\label{001}
 (\p_t)^2 \delta r - \frac{\cos^2 \theta}{ \sin^2 \psi}  \nabla^2 \delta r   = \frac{8 a_2}{a_0} \sin \theta \delta r +  \frac{8 a_2 }{ a_0}  \delta r   -  \frac{  16 a_0 + 20 a_2  }{5  a_0} \cos^2 \theta  \delta r \\
 +  \frac{ 4  }{5} \cos^2 \theta   \sin^2 \omega  \delta r 
\end{multline}
where $a_0 \approx 0.71805$, $a_2 = - (3 \times 6^{1/3})^{-1} $ are the warping constants of the KS throat (\ref{A44}) and $\nabla^2$ is the normalised Laplacian, i.e. $\nabla^2 = (\p_\omega)^2 + 1/\sin^2 \omega (\p_\varphi)^2 + \cot \omega \p_\omega$. 

\end{enumerate}

Before continuing, let us note an interesting fact about the $r$ extrinsic equation. If one follows the details in paragraph \ref{extrinsic}, it can be easily seen that the term 
\be
\frac{8 a_2 }{ a_0}  \sin \theta \delta r
\ee
is the $\tilde{F}_5$ electromagnetic force term while the terms
\be
\label{asdf}
\frac{8 a_2 }{ a_0}  \delta r  -  \frac{  16 a_0 + 20 a_2  }{5  a_0} \cos^2 \theta  \delta r  +  \frac{ 4  }{5} \cos^2 \theta   \sin^2 \omega  \delta r
\ee
are the gravitational force terms coming from the warping of the throat. The direction of the electromagnetic force term depends on the sign of the D3 brane charge carried by the KPV state $\mathcal{Q}_3 \sim C r_h^2 \sin \theta$. As KPV is a polarised state of anti-D3 branes, one might naively expect that this force is always attractive. However, this is not the case. The reason is because, in a fluxed setting, the D3 Page charge (\ref{71}) and the D3 brane charge (\ref{0010}) are not necessarily the same. In particular, for a range of $p/M$ near $p_{crit}$, the $\mathcal{Q}_3$ brane charge flips sign and, consequently, the electromagnetic force becomes repulsive. This effect can also be seen with the KP (Klebanov-Pufu) configuration \cite{Klebanov:2010qs} of anti-M2 branes at the tip of the CGLP (Cvetic-Gibbons-Lu-Pope) throat \cite{Cvetic:2000db}. Even though not explicitly stated, from the blackfold treatment of the KP state in \cite{M2M5brane}, one can easily infer the effect mentioned. 

\subsection{Stability analysis}
Immediately from the blackfold perturbation equations above, we see that the $\delta r$ variation decouples from other variations and is controlled only by equation (\ref{001}). This allows us to study separately stability of the non-radial perturbations and stability of the radial perturbations. For our convenience, before continuing, let us expand all our perturbations into momentum and spherical harmonic modes. We have
\begin{align}
\delta v^t &=  \int d \lambda \, e^{-i \lambda t} \sum^{\infty}_{l = 0} \sum^l_{m = -l} (S_{v^t})_l^m (\lambda) Y^m_l (\omega, \varphi) \\
\delta w^t &=  \int d \lambda \, e^{-i \lambda t} \sum^{\infty}_{l = 0} \sum^l_{m = -l} (S_{w^t})_l^m (\lambda) Y^m_l (\omega, \varphi) \\
\delta \tan \theta &=  \int d \lambda \, e^{-i \lambda t} \sum^{\infty}_{l = 0} \sum^l_{m = -l} (S_{\tan \theta})_l^m (\lambda) Y^m_l (\omega, \varphi) \\
\delta \psi &=  \int d \lambda \, e^{-i \lambda t} \sum^{\infty}_{l = 0} \sum^l_{m = -l} (S_\psi)_l^m (\lambda) Y^m_l (\omega, \varphi) \\
\delta r &=  \int d \lambda \, e^{-i \lambda t} \sum^{\infty}_{l = 0} \sum^l_{m = -l} (S_r)_l^m (\lambda) Y^m_l (\omega, \varphi)
\end{align}
where $Y^m_l(\omega, \varphi)$ are the standard spherical harmonics. Note that we do not write down the expansion for $\delta v^\omega$, $\delta w^\varphi$, and $\delta r_h$ because they can be expressed in term of other perturbations as shown in (\ref{5}), (\ref{6}), and (\ref{002}). 

\paragraph{Stability of non-radial perturbations}
Assuming $\lambda \neq 0$, expanding our perturbations in momentum and spherical harmonic modes, the $\omega$ intrinsic perturbation equation (\ref{9}) yields
\be
\sum^{\infty}_{l = 0} \sum^l_{m = -l} (S_{v^t})_l^m Y^m_l = - \frac{i \cot \theta \cos^2 \theta}{ \lambda \sqrt{M} b_0 \sin \psi}  \sum^{\infty}_{l = 0} \sum^l_{m = -l} (S_{\tan \theta})_l^m  \p_\omega  Y^m_l 
\ee
where $\lambda, \omega, \varphi$ dependence of $S^m_l (\lambda)$ and $Y^m_l (\omega, \varphi)$ have been subdued for syntactical simplicity. Similarly, from the $\varphi$ intrinsic perturbation (\ref{10}), we have
\be
\sum^{\infty}_{l = 0} \sum^l_{m = -l} (S_{w^t})_l^m Y^m_l = - \frac{i \cot \theta \cos^2 \theta}{ \lambda \sqrt{M} b_0 \sin \psi \sin \omega}   \sum^{\infty}_{l = 0} \sum^l_{m = -l} (S_{\tan \theta})_l^m  \p_\varphi  Y^m_l 
\ee
Let us note that satisfying the $\omega$ and $\varphi$ intrinsic perturbation equation automatically guarantee the satisfaction of the $J_2$ conservation equations (\ref{003})-(\ref{004}). Turning our attention to the $t$ intrinsic perturbation equation (\ref{005}), making use of the expressions above along with the identity $\nabla^2 Y^m_l = - l ( l + 1) Y^m_l$, we can show that 
\be
(S_{\tan \theta})^m_l = - \frac{2  \lambda^2  \sin^2 \psi \left( \cot \psi \tan \theta + 1/ b_0^2 \right) }{\lambda^2 \sin^2 \psi  -  l ( l + 1) \cos^2 \theta  } (S_\psi)^m_l
\ee
Again, let us note that satisfying the $t$ intrinsic perturbation equation automatically guarantee the satisfaction of the $J_4$ conservation equation (\ref{007}) and the conservation of $\mathbb{Q}_3$ charge (\ref{100}). Plugging in the expression of $(S_{\tan \theta})^m_l$ in term of $(S_\psi)^m_l$ into the $\psi$ extrinsic perturbation equation (\ref{008}) yields a quadratic equation for $\lambda^2$
\be
\lambda^4 + b \lambda^2 + c = 0
\ee
where the constants $b$ and $c$ are given respectively by
\begin{align}
b &= - \frac{4}{b_0^2} \cos^2 \theta (\sin \theta +1) \left( \cot \psi \tan \theta + \frac{1}{b_0^2} \right)-2 \left(l^2+l-1\right)   \frac{\cos^2\theta}{\sin^2 \psi} \\
c &= (l-1) l (l+1) (l+2) \frac{\cos^4\theta}{\sin^4 \psi} 
\end{align}
Then, it trivially follows that 
\be
\lambda^2 = \frac{- b \pm \sqrt{b^2 - 4 c}}{2}
\ee
It's important to remember that, as declared in the ``Perturbation parameters'' paragraph \ref{PerPara}, $\psi$ and $\theta$ denote the values of the variables evaluated at the KPV configuration. This means, for any KPV configuration, we can write down explicitly the values of $b$ and $c$, thus, the value of $\lambda^2$.

\begin{figure}\centering
\includegraphics[width= 0.8 \textwidth]{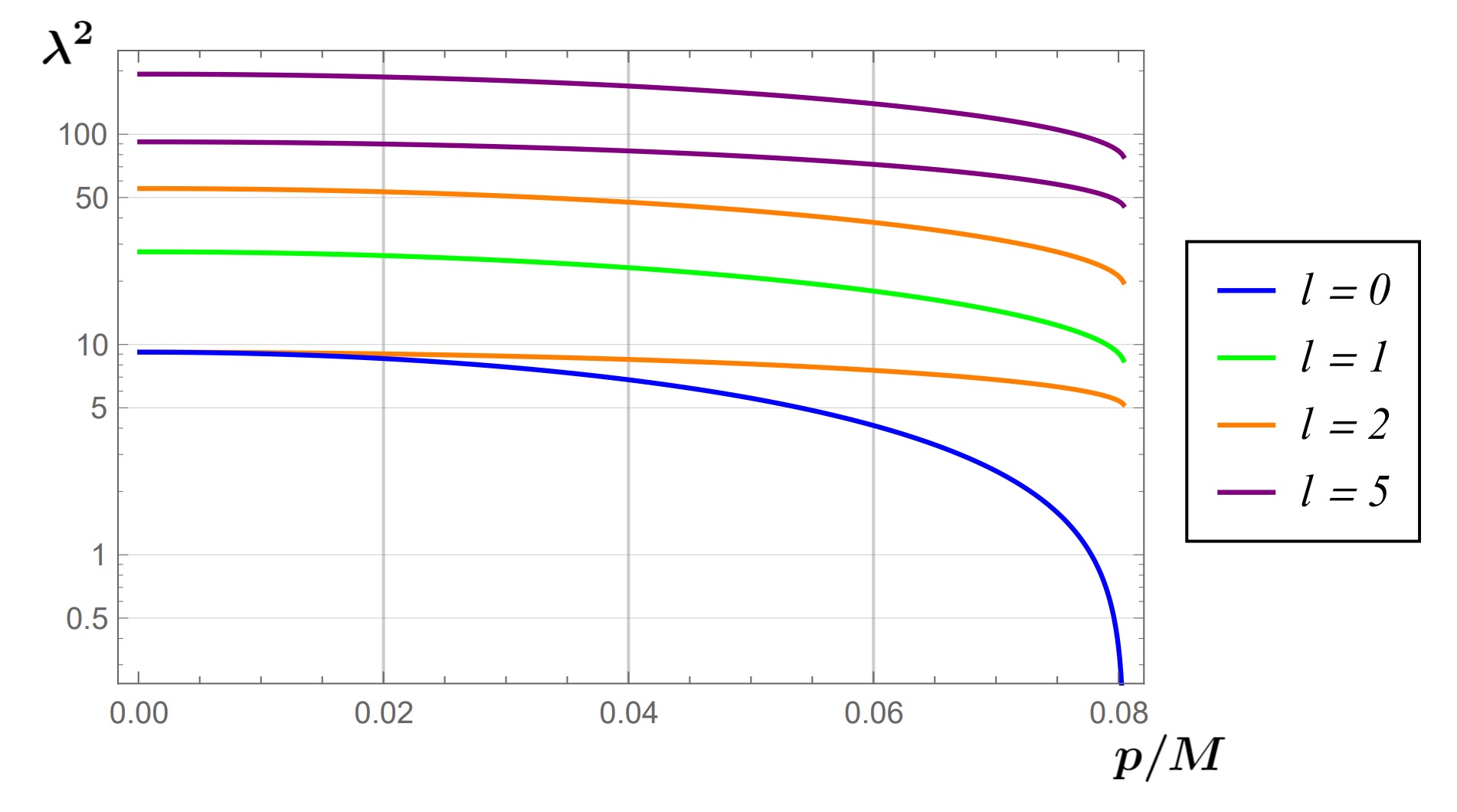}
\caption{\label{Fig1} Plot of $\lambda^2$ of non-radial perturbations against $p/M$. }
\end{figure}

It can be shown that $\lambda^2$ is positive for all KPV configurations. The case when $l = 0$ corresponds to having spherically homogeneous deformations around the KPV configuration and, as one would expect, it recreates the picture previously found. Including non-spherically homogeneous deformations does not change the statement regarding (meta)stability. In Figure \ref{Fig1}, we present the values of $\lambda^2$ for KPV configurations with $p/M \in (0, p_{crit})$ for $l$ equals $0$, $1$, $2$, and $5$.

Before continuing, let us ask the question: what happens if $\lambda = 0$? If $\lambda = 0$, the conservation of $\mathbb{Q}_3$ charge (\ref{100}) and the $\psi$ extrinsic perturbation equation (\ref{008}) both provide constraints on the $Y^0_0$ spherical harmonics mode of $\delta \tan \theta$ and $\delta \psi$. These conditions can only be simultaneously satisfied when
\be
\label{009}
\frac{1}{ \sin^2 \psi}   - \frac{2}{b_0^2}   \left( 1 + \sin \theta \right) \left( \tan \theta \cot \psi + \frac{1}{b_0^2} \right) = 0
\ee
Recall that the KPV states exist when the parameter $p/M$ is in the range $p/M \in (0,p_{crit})$  where $p_{crit} \approx 0.080488$. As one can easily checked, equation (\ref{009}) cannot be satisfied with any KPV states strictly in the regime $p/M \in (0,p_{crit})$. It is only satisfied when $p/M = p_{crit}$ as one would expect.

\paragraph{Stability of radial perturbations} 
Turning our attention to radial perturbations, expanding $\delta r$ in equation (\ref{001}) into momentum and spherical harmonic modes yields
\begin{multline}
\label{avv}
- \lambda^2 \sum^{\infty}_{l = 0} \sum^l_{m = -l} (S_r)_l^m  Y^m_l  + \frac{\cos^2 \theta}{\sin^2 \psi}   \sum^{\infty}_{l = 0} \sum^l_{m = -l} (S_r)_l^m  l (l + 1)Y^m_l    \\
= \Bigg( \frac{8 a_2}{a_0} \sin \theta  +  \frac{8 a_2 }{ a_0} -  \frac{16 a_0 + 20 a_2}{5  a_0} \cos^2 \theta + \frac{8}{15} \cos^2 \theta \Bigg) \sum^{\infty}_{l = 0} \sum^l_{m = -l} (S_r)_l^m  Y^m_l \\
- \frac{ 16}{15} \sqrt{\frac{\pi}{5}} \cos^2 \theta     \sum^{\infty}_{l = 0} \sum^l_{m = -l} (S_r)_l^m  Y^0_2 \,  Y^m_l 
\end{multline}
where we have used 
\be
\sin^2 \omega = \frac{2}{3} - \frac{4}{3} \sqrt{\frac{\pi}{5}} Y^0_2 
\ee
Considering spherical harmonic modes $Y^m_l$, we note that even though equation (\ref{avv}) doesn't mix $m$ modes, because of the $Y^0_2 \, Y^m_l$ contraction in the last term, $l$ modes are coupled and have to be studied together. Recall that the contraction of spherical harmonics with the $Y^0_2$ mode can be expressed as a sum of harmonics
\be
Y^0_2 \, Y^m_l = \sqrt{\frac{5 (2l +1)}{4 \pi}} \sum_{l_3} (-1)^{m} \sqrt{2 l_3 + 1} \left( \begin{matrix}
2  &  \ l & l_3 \\
0 & \ m & - m
\end{matrix}\right) \left( \begin{matrix}
2  & \ l & \ l_3 \\
0 & \ 0 & \ 0
\end{matrix}\right) Y_{l_3}^{m} 
\ee
where $\left( \begin{matrix}
2  & \ l & l_3 \\
0 & \ m & - m
\end{matrix}\right)$
and
 $\left( \begin{matrix}
2  & \ l & \ l_3 \\
0 & \ 0 & \ 0
\end{matrix}\right)$
are the Wigner 3j-symbols, which vanish unless $|l - 2| \leq l_3 \leq l + 2$. By writing down the condition for each individual $l$ mode, equation (\ref{avv}) can be expressed as a set of linear equations of $(S_r)^m_l$.  

As $m$ modes decoupled, let us discuss in details the spherical harmonic modes with $m = 0$. The associated matrix of the linear system of $(S_r)^0_l$ is given by
\be
\begin{pmatrix} \mathbb{A} \ \vline \ 0
\end{pmatrix}
=
\begin{pmatrix}
\lambda^2 + d & 0 &  \frac{ - 8 \cos^2 \theta}{15 \sqrt{5}} & \dots &\vline & \ 0\\ 
0 & \lambda^2 + d - \frac{2 \cos^2 \theta}{\sin^2 \psi} - \frac{16}{75} \cos^2 \theta & 0 & \dots &\vline & \ 0\\ 
\frac{ - 8 \cos^2 \theta}{15 \sqrt{5}} & 0 & \lambda^2 + d - \frac{6 \cos^2 \theta}{\sin^2 \psi} - \frac{16}{105} \cos^2 \theta & \dots &\vline & \ 0\\

\vdots & \vdots  & \vdots  & \ddots &\vline & \ \vdots
\end{pmatrix} 
\ee
where, for convenience, we have defined a constant $d$ as
\be
d = \frac{8 a_2}{a_0} \sin \theta  +  \frac{8 a_2 }{ a_0} -  \frac{16 a_0 + 20 a_2}{5  a_0} \cos^2 \theta + \frac{8}{15} \cos^2 \theta 
\ee
The system of linear equations is only satisfied when the determinant of the associated matrix vanishes, i.e. $\det \mathbb{A} = 0$. Even though
$\mathbb{A}$ is not diagonal, as the contribution of the off-diagonal terms to the determinant of $\mathbb{A}$ is numerically much smaller than that of the diagonals, the determinant of $\mathbb{A}$ can be well-approximated by the product of the diagonal terms. With this approximation, it's trivial that $\lambda^2$ is always positive. Let us mention also that cases of $m \neq 0$ can be treated the same way and yield a similar conclusion. 

\begin{figure}\centering
\includegraphics[width= 0.8 \textwidth]{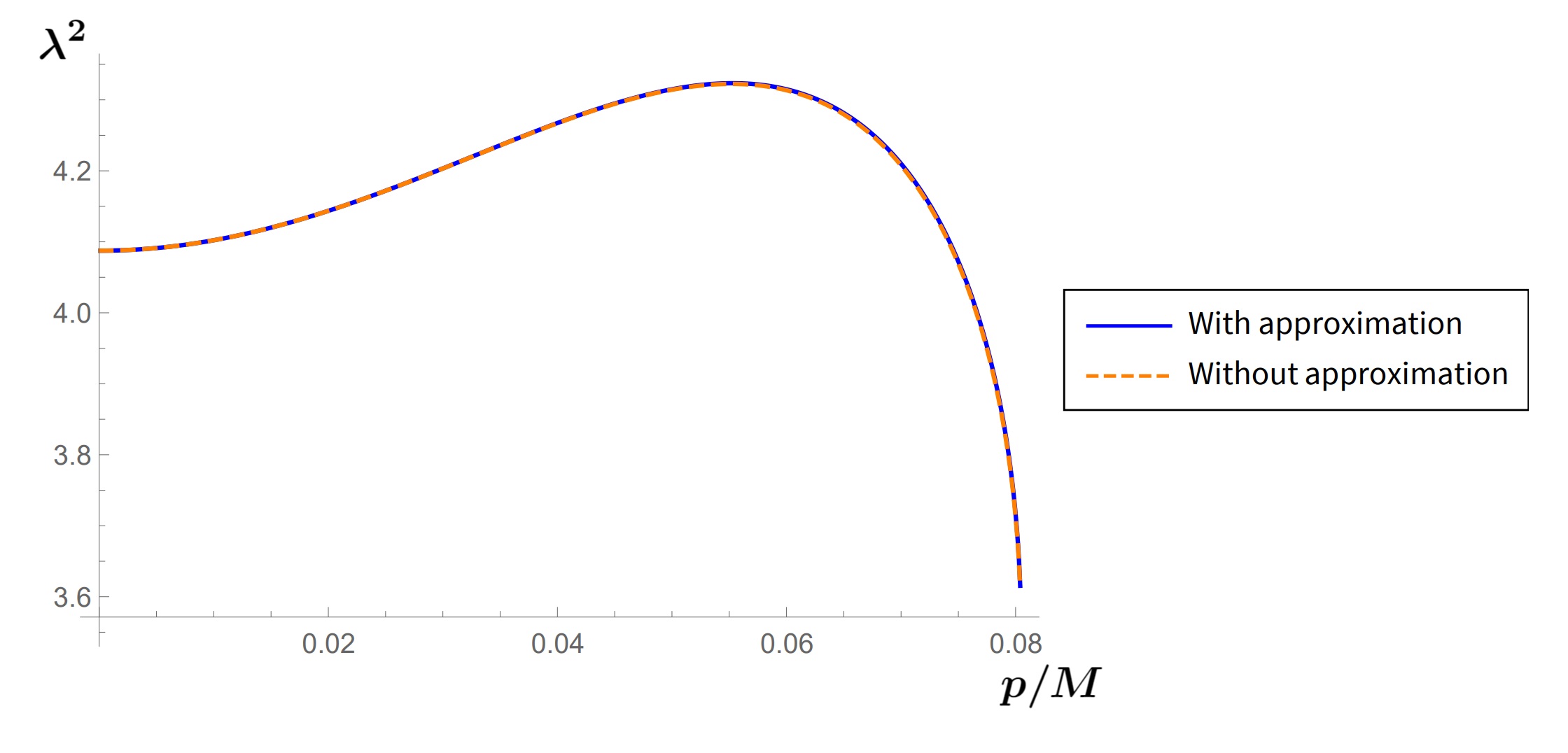}
\caption{\label{Fig2} Plot of $\lambda^2$ of radial perturbations against $p/M$. }
\end{figure}

In Figure \ref{Fig2}, we plotted the smallest $\lambda^2$ root computed both with the diagonal approximation\footnote{Practically, this is a plot of $\lambda^2 = -d$} and without the diagonal approximation, truncating $\mathbb{A}$ to be of order $21 \times 21$. From the plot, it can easily be seen that the off-diagonal corrections are indeed very minimal and don't affect the underlying physics of the system. Lastly, let us note that the dip in $\lambda^2$ near $p_{crit}$ is because of the effect mentioned in the discussion below equation (\ref{asdf}) where the $\mathcal{Q}_3$ charge flips sign and the electromagnetic force becomes repulsive. Nevertheless, as demonstrated here, this electromagnetic repulsion is outweighed by gravitational attraction.

\acknowledgments
We would like to especially thank Jay Armas, Vasilis Niarchos, Niels Obers, and Thomas Van Riet for useful discussions, suggestions and collaboration on a related project \cite{Armas:2018rsy}. A special thank to Vasilis Niarchos for guidance on multiple aspects of this paper.  

\begin{appendices}
\section{Klebanov-Strassler throat}\label{KS}
The Klebanov-Strassler (KS) throat is a 10-dimensional type IIB supergravity solution. The throat involves a 6 dimensional deformed conifold, a 4 dimensional Minkowskian space, and non-trial $F_3, F_5, H_3$ fluxes, which in turn induce warping effects on the flat space and the conifold. In this appendix, we shall discuss aspects of the KS throat that are immediately relevant for us. For a complete discussion of the KS throat, we refer the readers to the original paper \cite{Klebanov:2000hb} or the review \cite{Herzog:2001xk}.

\subsection{The 6-dimensional deformed conifold} The 6 dimensional deformed conifold of the KS solution is given by the equation
\be
\sum^4_{i = 1} z_i^2 = \varepsilon^2
\ee
where $z_i$ are complex numbers and $\varepsilon$ characterises the degree of deformation, i.e. if $\varepsilon = 0$, we have a normal cone. In order to obtain a parametrisation of the space, a clever trick one can do is to define the matrix
\be
W = \left( \begin{matrix}
z_3 + i z_4 & z_1 - i z_2 \\
z_1 + i z_2 & - z_3 + i z_4 
\end{matrix}\right)
\ee
then the defining equation becomes
\be
\det W = - \varepsilon^2
\ee
It's easy to see that 
\be
W_0 = \left( \begin{matrix}
0 & \varepsilon e^{\tau/2} \\
\varepsilon e^{- \tau/2} & 0
\end{matrix}\right)
\ee
is one possible solution. Furthermore, if we define two $SU(2)$ matrices $L_j$ with $j = 1,2$ then 
\be
W = L_1 . W_0 . L_2^{\dagger}
\ee
also satisfies the equation $\det W = - \varepsilon^2$. As argued in \cite{Minasian:1999tt}, the metric of the deformed conifold is then given by 
\be
\label{a}
ds^2 = \mathcal{F} tr \left( d W^{\dagger} d W \right) + \mathcal{G} | tr (W^\dagger d W) |^2
\ee
where
\begin{gather}
\mathcal{F} (\tau) =  \frac{(\sinh 2 \tau  - 2 \tau)^{1/3}}{2 \times 2^{1/3} \times   \varepsilon^{2/3} \sinh \tau} \\
\mathcal{G}(\tau) = \frac{2 - 3 \coth^2 \tau + 3 \tau (\cosh \tau/ \sinh^3 \tau) }{12  \times   \varepsilon^{8/3} (\cosh \tau \sinh \tau - \tau)^{2/3}}
\end{gather}

\paragraph{Angular parametrisation of the deformed conifold}
One can parametrise the $L_j$ matrices using Euler angles as
\be
L_j = \left( \begin{matrix}
\cos \frac{\theta_j}{2} e^{i (\psi_j + \phi_j)/2} & - \sin \frac{\theta_j}{2} e^{-i (\psi_j - \phi_j)/2} \\
\sin \frac{\theta_j}{2} e^{i (\psi_j - \phi_j)/2} & \cos \frac{\theta_j}{2} e^{-i (\psi_j + \phi_j)/2} 
\end{matrix}\right)
\ee
with $(\psi_j, \phi_j)$ range from $0$ to $ 2 \pi$ and $\theta$ ranges from $0$ to $\pi$. Plugging the parametrised expression of $W = L_1 . W_0 . L_2^{\dagger}$ into (\ref{a}) yields the metric of the deformed conifold written in angular coordinates $\psi_j, \theta_j, \phi_j$. As the coordinates $\psi_1$ and $\psi_2$ only appear in $W$ as $\psi_1 + \psi_2$, we can define a new coordinate $\psi = \psi_1 + \psi_2$. The deformed conifold metric in these coordinates is then given by
\begin{equation}
ds_6^2 = \frac{1}{2} \varepsilon^{4/3} K(\tau) \Bigg[ \frac{1}{3 K^3 (\tau) } (d \tau^2 + (g^5)^2 ) + \cosh^2 \left ( \frac{\tau}{2} \right) [(g^3)^2 + (g^4)^2]  + \sinh^2 \left( \frac{\tau}{2} \right) [(g^1)^2 + (g^2)^2  ] 
\Bigg]
\end{equation}
where the function $K(\tau)$ is given by
\be
K (\tau) = \frac{(\sinh 2 \tau  - 2 \tau)^{1/3}}{2^{1/3} \sinh \tau} 
\ee
and the $g^i$ forms are given by
\begin{align}
g^1 &=  \frac{- \sin \theta_1 d \phi_1 - \cos \psi \sin \theta_2 d \phi_2 + \sin \psi d \theta_2}{\sqrt{2}} \\
g^2 &=  \frac{d \theta_1 - \sin \psi \sin \theta_2 d \phi_2 - \cos \psi d \theta_2}{\sqrt{2}} \\
g^{3} &= \frac{- \sin \theta_1 d \phi_1 + \cos \psi \sin \theta_2 d \phi_2 - \sin \psi d \theta_2}{\sqrt{2}} \\
g^{4} &= \frac{d \theta_1 + \sin \psi \sin \theta_2 d \phi_2 + \cos \psi d \theta_2}{\sqrt{2}}\\
g^5 &= d \psi + \cos \theta_1 d \phi_1 + \cos \theta_2 d \phi_2
\end{align}
where $\psi$ is a special angular coordinate going from $0$ to $4 \pi$ while $(\theta_j, \phi_j)$ are the standard $S^2$ spherical coordinate going from $0$ to $\pi$ and $0$ to $2 \pi$ respectively.

Let us note further that, as argued in \cite{Minasian:1999tt}, the metric 
\be
ds^2 = \frac{1}{2} (g^5)^2 + (g^4)^2 + (g^3)^2 
\ee
and \be
ds^2 = (g^1)^2 + (g^2)^2
\ee
are the metric of respectively the standard $S^3$ sphere with radius $\sqrt{2}$ and the standard $S^2$ sphere with radius $\sqrt{2}$. 

\subsection{Klebanov-Strassler throat near the apex in Euler angles}
For the leading order stability analysis of the KPV state, we are only interested in the description of the KS throat near the apex. From the full description of the throat, we expand the metric and gauge fields in $\tau$ and keep only the relevant terms. To be more specific, we keep in the metric and gauge fields terms of the required order such that the profile of metric and fields solve the Supergravity equations to first order in $\tau$. For convenience, let us also set\footnote{Setting $g_s = 1$ is possible because the KS solution a has constant dilaton.} $g_s = 1$ and $\alpha' = 1$ in all our discussions of the KS throat. 

The KS metric near the apex is approximated by
\begin{multline}
ds_{10}^2 = A_1(\tau) \left( - (dx^0)^2 + ( d x^1)^2+ (d x^2)^2 + (dx^3)^2 \right) + A_2 (\tau) \left( d(\tau )^2 + (g^5)^2 \right) \\
+ A_3 (\tau) \left( (g^3)^2+ (g^4)^2 \right) + A_4 (\tau) \left( (g^1)^2 + (g^2)^2 \right)
\end{multline}
where
\be
A_1 (\tau) = \frac{\epsilon^{4/3}}{2^{1/3} (a_0)^{1/2} M} - \frac{ a_2 \, \tau^2 \, \epsilon^{4/3}}{2 \times 2^{1/3} (a_0)^{3/2} M} + \frac{3 \, (a_2)^2 \, \tau^4 \, \epsilon^{4/3}}{8 \times 2^{1/3} (a_0)^{5/2} M} -\frac{a_4 \, \tau^4 \, \epsilon^{4/3}}{2 \times 2^{1/3} (a_0)^{3/2} M} 
\ee
\begin{multline}
A_2 (\tau) = \frac{(a_0)^{1/2} M}{2 \times 6^{1/3}} + \frac{(a_0)^{1/2} M \, \tau^2}{10 \times 6^{1/3}} + \frac{a_2 \, M \, \tau^2}{4 \times 6^{1/3} (a_0)^{1/2}} - \frac{ (a_2)^2 M \, \tau^4}{16 \times 6^{1/3} (a_0)^{3/2}} \\ 
+\frac{(a_0)^{1/2} M \, \tau^4}{210 \times 6^{1/3}}  +\frac{ a_4 \, M \, \tau^4}{4 \times 6^{1/3} (a_0)^{1/2}} + \frac{a_2 \, M \, \tau^4}{20 \times 6^{1/3} (a_0)^{1/2}}
\end{multline}
\begin{multline}
A_3 (\tau) = \frac{(a_0)^{1/2} M}{6^{1/3}} + \frac{3^{2/3} (a_0)^{1/2} M \, \tau^2}{20 \times 2^{1/3}} +  \frac{a_2 \, M \, \tau^2}{2 \times 6^{1/3} (a_0)^{1/2}} + \frac{a_4 \, M \, \tau^4}{2 \times 6^{1/3} (a_0)^{1/2}}  \\
+ \frac{17 \, (a_0)^{1/2} M \, \tau^4}{2800 \times 6^{1/3}}   - \frac{(a_2)^2 M \, \tau ^4}{8 \times 6^{1/3} (a_0)^{3/2}} + \frac{3^{2/3} a_2 \, M \tau^4}{40 \times 2^{1/3} (a_0)^{1/2}} 
\end{multline}
\begin{multline}
A_4 (\tau)  = \frac{(a_0)^{1/2} M \, \tau ^2}{4 \times 6^{1/3}}  - \frac{(a_0)^{1/2} M \, \tau^4}{240 \times 6^{1/3}} + \frac{a_2 \, M \, \tau ^4}{8 \times 6^{1/3} (a_0)^{1/2}} + \frac{a_4 \, M \, \tau^6}{8 \times 6^{1/3} (a_0)^{1/2}} \\
- \frac{(a_2)^2 M \, \tau^6}{32 \times 6^{1/3} (a_0)^{3/2}} - \frac{ a_2 \, M \, \tau ^6}{480 \times 6^{1/3} (a_0)^{1/2}} +\frac{59 \, (a_0)^{1/2} \, M \,\tau ^6}{50400 \times 6^{1/3}}
\end{multline}
with the constants $a_0 \approx 0.71805$, $a_2 = - (3 \times 6^{1/3})^{-1} $, and $a_4 = (18 \times 6^{1/3})^{-1}$.

The KS fluxes near the apex are approximated by\footnote{As our convention of the Hodge star operator is different from that of \cite{Klebanov:2000hb}, our description of $H_3$ and $\tilde{F}_5$ have different signs from those of \cite{Klebanov:2000hb}.}
\begin{multline}
H_3 = - \frac{M}{2} \Bigg( \left( \frac{\tau^2}{4} - \frac{\tau^4}{16} \right) d \tau \wedge g^1 \wedge g^2 + \left( \frac{1}{3} + \frac{\tau^2}{60} + \frac{\tau^4 }{1008} \right) d \tau \wedge g^3 \wedge g^4  \\
+ \left( \frac{\tau}{6} - \frac{7}{180} \tau^3 \right) g^5  \wedge (g^1 \wedge g^3 + g^2 \wedge g^4) \Bigg)
\end{multline}
\begin{multline}
H_7 = - \frac{\epsilon^{8/3}}{2 \times 2^{2/3} a_0 M} \Bigg( \left(1 - \frac{\tau^2}{12} - \frac{a_2 \tau^2 }{a_0}\right) d x^0 \wedge ... \wedge dx^3 \wedge g^3 \wedge g^4 \wedge g^5 \\
+ \frac{\tau}{6} d x^0 \wedge ... d x^3 \wedge d \tau \wedge \left( g^1 \wedge g^3 + g^2 \wedge g^4 \right) + \frac{\tau^2}{12} d x^0 \wedge ... \wedge d x^3 \wedge g^1 \wedge g^2 \wedge g^5 \Bigg)
\end{multline}

\begin{multline}
F_3 = \frac{M}{2} \Bigg( \left( 1 - \frac{\tau^2}{12} + \frac{7 \, \tau^4}{720} \right) g^5 \wedge g^3 \wedge g^4 + \left( \frac{\tau^2}{12}  - \frac{7 \, \tau^4}{720} \right)g^5 \wedge g^1 \wedge g^2 \\
+ \left ( \frac{\tau}{6} - \frac{7 \, \tau^3 }{180} \right) d \tau \wedge (g^1 \wedge g^3 + g^2 \wedge g^4 ) \Bigg)
\end{multline}
\be
F_5 = \frac{\epsilon^{8/3}}{M^2} \left( \frac{\tau}{3 \times 3^{1/3} a_0^2 } - \frac{\tau^3}{9 \times 3^{1/3} \, a_0^2} - \frac{2 \, a_2 \, \tau^3}{3 \times 3^{1/3} \, a_0^3} \right) dx^0 \wedge dx^1 \wedge dx^2 \wedge dx^3 \wedge d \tau
\ee
\begin{multline}
\label{201}
\tilde{F}_5 = \frac{\epsilon^{8/3}}{M^2} \left( \frac{\tau}{3 \times 3^{1/3} a_0^2 } - \frac{\tau^3}{9 \times 3^{1/3} \, a_0^2} - \frac{2 \, a_2 \, \tau^3}{3 \times 3^{1/3} \, a_0^3} \right) dx^0 \wedge dx^1 \wedge dx^2 \wedge dx^3 \wedge d \tau\\
-  \left( \frac{M^2 \, \tau^3}{36} \right) g^1 \wedge g^2 \wedge g^3 \wedge g^4 \wedge g^5  
\end{multline}

\subsection{Klebanov-Strassler metric near the apex in adapted coordinates}
The description of the KS throat near the apex above is in the angular coordinates $x^0$, $x^1$, $x^2$, $x^3$,
$\tau$, $\psi$, $\theta_1$, $\phi_1$, $\theta_2$, $\phi_2$ as presented in the original paper of Klebanov and Strassler. However, for our purpose, it proves useful to express the KS metric near the apex in adapted coordinates $t$, $x^1$, $x^2$, $x^3$, $r$, $\psi$, $\omega$, $\varphi$, $\tilde{\omega}$, $\tilde{\varphi}$ as used in the rest of the paper\footnote{Note that the duplicate coordinates $x^1,x^2,x^3$, and $\psi$ of the two coordinates system are different. We decided not to change them to be consistent with the literature.}.

One might also wish to write the fluxes in term of the adapted coordinates. But, as the fluxes enter the blackfold equations only when coupled to the anti-D3-NS5 currents, only some components are relevant. As a result, we shall not attempt to transform the full description of the fluxes to the adapted coordinates but only the relevant components when needed.  

The Minkowskian coordinates $x^0$, $x^1$, $x^2$, $x^3$ and the radial coordinates $\tau$ of the angular coordinate system are respectively, up to some scaling, equivalent to the coordinates $t$, $x^1$, $x^2$, $x^3$, and $r$ used in the rest of the paper. In particular, one can transform from one to the other as
\begin{align}
\label{1002}
x_0 &\rightarrow \frac{\sqrt{2} \sqrt{a_0} M}{3^{1/6} \times  \epsilon^{2/3}} \ t \\
x_i &\rightarrow \frac{\sqrt{2} \sqrt{a_0} M}{3^{1/6} \times  \epsilon^{2/3}} \ x_i \\
\label{1003}
\tau &\rightarrow 2 \ r 
\end{align}
Let us turn to the base of the conifold, which originally was expressed using Euler angles  $(\psi$, $\theta_1$, $\phi_1$, $\theta_2$, $\phi_2 )$, and attempt to parametrise it using  the spherical coordinates $( \psi, \omega, \varphi, \tilde{\omega}, \tilde{\varphi})$.

\paragraph{Spherical parametrisation of the deformed conifold}
For our analysis, it's most convenient to parametrise both the $S^3$ at the tip and the transverse $S^2$ using spherical coordinates, i.e. $(\psi, \omega, \varphi)$ and $(\tilde{\omega}, \tilde{\varphi})$ respectively. To do this, we shall apply the same parametrisation process as before but with an emphasis on identifying the 3 parameters of the tip $S^3$ and incorporate the remaining 2 parameters as we go up the throat. Recall from (\ref{a}) that the metric of the deformed conifold is given by
\be
\label{G}
ds^2 = \mathcal{F} tr \left( d W^{\dagger} d W \right) + \mathcal{G} | tr (W^\dagger d W) |^2
\ee
where 
\be
W = L_1 . W_0 . L_2^{\dagger}
\ee
with
\be
W_0 = \left( \begin{matrix}
0 & \varepsilon e^{\tau/2} \\
\varepsilon e^{- \tau/2} & 0
\end{matrix}\right)
\ee
and $L_j$ with $j=1,2$ are two $SU(2)$ matrices. As noted before that the coordinates $\psi_1$ and $\psi_2$ only appear in $W$ as $\psi_1 + \psi_2$, so instead of relabelling the final result, we parametrise $L_2$ with only two variables $(\theta_2, \phi_2)$
\be
L_2 = \left( \begin{matrix}
\cos \frac{\theta_2}{2} e^{i \phi_2/2} & - \sin \frac{\theta_2}{2} e^{i \phi_2/2} \\
\sin \frac{\theta_2}{2} e^{- i \phi_2/2} & \cos \frac{\theta_2}{2} e^{-i \phi_2/2} 
\end{matrix}\right)
\ee 

Expanding $W_0$ in $\tau$, we have 
\be
\label{1000}
W_0 = \varepsilon f(\tau) \sigma_1 + \varepsilon  g(\tau) \sigma_2
\ee
where
\begin{align}
&\sigma_{1} = \left( \begin{matrix}
0 & 1 \\
1 & 0 
\end{matrix}\right) & & \sigma_2 = \left( \begin{matrix}
0 & 1 \\
- 1 & 0 
\end{matrix}\right)
\end{align}
and
\begin{align}
f(\tau) &=  1 + \frac{\tau^2}{8} + \frac{\tau^4}{384} + \mathcal{O} \left( \tau^6 \right)  &
g(\tau) &=  \frac{\tau}{2} + \frac{\tau^3}{48} + \mathcal{O} \left( \tau^5 \right)
\end{align}
Thus, we have
\begin{align}
W &= L_1 . \big( \varepsilon f(\tau) \sigma_1 + \varepsilon  g(\tau) \sigma_2 \big) . L_2^\dagger \\
&= \varepsilon f(\tau) L + \varepsilon g(\tau) L . \hat{L}   
\end{align}
where $L \equiv L_1 . \sigma_1 . L_2^\dagger$ and $\hat{L} \equiv L_2 . (\sigma_1)^{-1} . \sigma_2 . L_2^\dagger$.

As $L$ is an unitary complex matrix with $\det L = - 1$, we can parametrise $L$ using spherical coordinates as\footnote{To obtain the deformed conifold metric, it's algebraically simpler to write the matrix $L$ in Hopf coordinates first, carry out the necessary computations, then transform Hopf to spherical. Nevertheless, the final answers are the same.}
\be
L = \left( \begin{matrix}
- \sin \psi \sin \omega \cos \varphi + i \sin \psi \sin \omega \sin \varphi & \cos \psi - i \sin \psi \cos \omega \\
\cos \psi + i \sin \psi \cos \omega & \sin \psi \sin \omega \cos \varphi + i \sin \psi \sin \omega \sin \varphi 
\end{matrix}\right)
\ee
On the other hand, the parametrisation of $\hat{L}$ comes directly from the parametrisation of $L_2$. We have
\be
\hat{L} = \left( \begin{matrix}
- \cos \theta_2 & - e^{i \phi_2} \sin \theta_2 \\
- e^{- i \phi_2} \sin \theta_2 & \cos \theta_2
\end{matrix}\right)
\ee
Plugging the spherically parametrised $W$ into (\ref{G}), we obtain the metric of the deformed conifold in spherical coordinates.

\paragraph{Klebanov-Strassler metric near the apex in adapted coordinates} Recall from \cite{Klebanov:2000hb}, the KS metric is given by
\be
ds_{10}^2 = h^{-1/2} (\tau) \left( - d x_0^2 + d x_1^2 + dx_2^2 + dx_3^2 \right) + h^{1/2} (\tau) ds_6^2
\ee
where $ds_6^2$ is the metric of the deformed conifold and the $h(\tau)$ is the warping effects induced by the non-trivial fluxes:
\begin{align}
\label{A44}
h (\tau) &= M^2 \, 2^{2/3} \epsilon^{-8/3} \int_\tau^\infty dx \frac{x \coth x - 1}{\sinh^2 x} (\sinh 2 x - 2 x)^{1/3}\\
&= M^2 2^{2/3} \epsilon^{-8/3} \ ( a_0 + a_2 \tau^2  + a_4 \tau^4)  + \mathcal{O} (\tau^6)
\end{align}
where, as written down earlier, $a_0 \approx 0.71805$, $a_2 = - (3 \times 6^{1/3})^{-1} $, and $a_4 = (18 \times 6^{1/3})^{-1}$.

Substituting in the spherically parametrised deformed conifold metric, applying the coordinate transformations (\ref{1002} - \ref{1003}), relabelling $\theta_2 \rightarrow \tilde{\omega}$ and $\phi_2 \rightarrow \tilde{\varphi}$, and restricting our attention to some leading orders of $r$, we obtain the expression of the KS metric near the apex in our desired adapted coordinates. However, as the expression is long and ugly, we shall not write it explicitly here. Instead, we shall only write down components/properties that are immediately relevant for us.

Firstly, as you would as expect, if we subdue terms of order $r^2$ or higher in all but the $(\tilde{\omega}, \tilde{\varphi})$ directions, we recover the metric in (\ref{200}):
\begin{multline}
g_{\mu \nu} d x^\mu d x^\nu = M b_0^2  \Big ( - dt^2 + (dx^1)^2 + (dx^2)^2 + (dx^3)^2  + dr^2 \\
+ d \psi ^2 + \sin^2 \psi \left( d \omega^2 + \sin^2 \omega d \varphi^2 \right)  + r^2 (d \tilde{\omega}^2 +  \sin^2 \tilde{\omega} d \tilde{\varphi}^2) \Big)
\end{multline}
where $b_0^2= \frac{2^{2/3} \sqrt{a_0}}{3^{1/3}} \approx 0.93266$.

Secondly, as they will be relevant for our stability analysis, we note the following derivatives
\begin{align}
&\p_r^2 g_{tt}\Big|_{r = \tilde{\omega} = \tilde{\varphi} = 0}  = \frac{4 \times 2^{2/3} a_2 M}{3^{1/3} \sqrt{a_0}}  &  &\p_r^2 g_{x^i x^i}\Big|_{r = \tilde{\omega} = \tilde{\varphi} = 0} =  - \frac{4 \times 2^{2/3} a_2 M}{3^{1/3} \sqrt{a_0}}
\end{align}
\be
\p_r^2 g_{\omega \omega} \Big|_{r = \tilde{\omega} = \tilde{\varphi} = 0}  = \frac{ 4 \times 2^{2/3} M }{5 \times 3^{1/3} \sqrt{a_0}} \sin ^2 \psi  \Big(  4 a_0 + 5 a_2 - 2 a_0 \cos^2 \psi \sin^2 \omega \Big)
\ee
\be
\p_r^2 g_{\varphi \varphi} \Big|_{r = \tilde{\omega} = \tilde{\varphi} = 0}  =
\frac{4 \times 2^{2/3} M }{5 \times 3^{1/3} \sqrt{a_0}} \sin ^2\psi \sin ^2\omega  \left(4 a_0+5 a_2 - 2 a_0 \sin^2 \psi \sin ^2 \omega \right)
\ee
with $a_0 \approx 0.71805$ and $a_2 = - (3 \times 6^{1/3})^{-1} $.



\section{D3-NS5 branes} \label{secA}
\subsection{D3-NS5 supergravity solution}\label{D3-NS5Sugra}
For the convenience of the readers, let us present here the known supergravity description of the D3-NS5 bound state as well as its thermodynamic data (see \cite{Harmark:1999rb, Emparan:2011hg} for detailed discussion). In the string frame, the metric is given by
\begin{equation}\label{3}
ds^2 = D^{-1/2} \left( -f dt^2 + D \left( (dx^1)^2 + (dx^2)^2 \right)+ \sum_{i=3}^5 (dx^i)^2 \right)+ H D^{-1/2} \left( f^{-1} dr^2 + r^2 d\Omega_3^2 \right)
\end{equation}
with
\begin{align}
f &= 1 - \frac{r^2_0}{r^2} , & D = \left( \sin^2 \theta H^{-1} + \cos^2 \theta \right) ^{-1}\\
 H &=1 + \frac{r_0^2 \sinh^2 \alpha}{r^2} 
\end{align}
where $d \Omega_3^2$ is the standard $S^3$ metric $d \Omega_3^2 = d \psi^2 + \sin^2 \psi \left( d \omega^2 + \sin^2 \omega d \varphi^2 \right)$. The dilaton field is given by
\be
e^{2 \phi} = H D^{-1}
\ee
and the gauge fields are given by
\begin{align} 
C_2 &= - \tan \theta (H^{-1} D - 1) \, d x^1 \wedge d x^2 \\
B_2 &= - 2 r_0^2 \sinh^2 \alpha \cos \theta \ \varphi \sin^2 \psi \sin \omega d \psi \wedge d \omega \\
\label{4}
C_4 &= (H^{-1} - 1) \sin \theta \, d t \wedge d x^3 \wedge d x^4 \wedge dx^5 + \frac{r^2}{r_0^2 \sinh^2 \alpha \cos^2 \theta} B_2 \wedge C_2
\end{align}
The thermodynamics of this solution are
\begin{align}
&\varepsilon = \frac{\Omega_3}{16\pi G} r_0^2 \left( 3 + 2 \sinh^2 \alpha \right) & &s = \frac{\Omega_3}{4G} r_0^3 \cosh\alpha & &\mathcal{T} = \frac{1}{2\pi r_0 \cosh\alpha}
\end{align}
\begin{align}
\label{0010}
&\Phi_3 = \sin\theta\, \tanh\alpha & & \mathcal{Q}_3 = \frac{\Omega_3}{8\pi G} r_0^2\, \sin\theta \, \sinh\alpha \, \cosh\alpha \\
&\Phi_5 = \cos\theta\, \tanh\alpha & & \mathcal{Q}_5 = \frac{\Omega_3}{8\pi G} r_0^2\, \cos\theta \, \sinh\alpha \, \cosh\alpha
\end{align}
where $\Omega_3=2\pi^2$ is the volume of the unit radius round $S^3$. And, the effective energy stress tensor is given by
\be
T_{ab} = \mathcal{T} s \left( u_a u_b - \frac{1}{n} \gamma_{ab} \right) - \sum_{q \, = \, 3, 5} \Phi_q \mathcal{Q}_q h_{ab}^{(q)} 
\ee
The extremal D3-NS5 solution can be obtained by taking the limit $r_0 \rightarrow 0, \alpha \rightarrow \infty$ in such a way that we can define a finite extremal horizon radius $r_h \equiv r_0 \sinh \alpha$. In fact, for the purpose of this paper, we shall only be interested in the D3-NS5 solution in the extremal limit. 

\subsection{Far-zone equivalent currents}
As discussed in \cite{Marolf:2000cb}, there are at least three sensible notions of charges in a supergravity theory. For the purpose of constructing equivalent currents, we shall be interested in something called the Maxwell charge. The key idea for the Maxwell charges is that the Chern-Simons terms in the equation of motion can be thought of as a source for the gauge field. For example, let us look at the equation of motion for the $C_4$ gauge field in type IIB supergravity: 
\be
d \star \tilde{F}_5 - H_3 \wedge F_3 = - 16 \pi G \star J_4
\ee
In this case, the Maxwell current is given by
\be
d \star \tilde{F}_5 = - 16 \pi G \star J_{4}^{Maxwell} = - 16 \pi G \star J_4 + H_3 \wedge F_3
\ee
where the sign and factors in front of $J_4^{Maxwell}$ is to make sure it is compatible with our conventions of $J_4$. The Maxwell charge can be computed easily from Gauss's law of the $\tilde{F}_5$ flux and, thus, can be interpreted as the monopole source that will reproduce the $\tilde{F}_5$ flux far away. 

Turning our attention to the case of D3-NS5 branes, we have the relevant forced Maxwell equations are
\begin{align}
d \star \tilde{F}_3 &= - 16 \pi G \star J_2^{Maxwell} \\
d \star \tilde{F}_5 &= -16 \pi G \star J_4^{Maxwell} \\
d \star H_7 &= 16 \pi G \star j_6^{Maxwell}  
\end{align}
We do not know the exact expressions of these Maxwell currents, however, we can mimic their effects far away by using Maxwell charges to construct a set of equivalent currents. Adopting the convention that $Q = \int \star J$, using the description of extremal D3-NS5 branes in (\ref{3})-(\ref{4}), we obtain the Maxwell charges
\begin{align}
Q_1^{Maxwell} &= Vol_4 \ C r_h^2 \sin \theta \cos \theta \\
\label{7}
Q_3^{Maxwell} &= Vol_2 \ C r_h^2 \sin \theta \\
Q_5^{Maxwell} &= - C r_h^2 \cos \theta
\end{align}
Requiring that they reproduce the same Maxwell charges at $r \rightarrow \infty$, our equivalent currents can now be easily constructed. These are\footnote{The equivalent currents are localised ($\delta$ function) currents in the full 10 dimensional picture.}
\begin{align}
J_2^{equiv} &= C r_h^2 \sin \theta \cos \theta \ v \wedge w \\
J_4^{equiv} &=   C r_h^2 \sin \theta \ * ( - v \wedge w) \\
j_6^{equiv} &= - C r_h^2 \cos \theta \ * (- 1) 
\end{align}
where $*$ is the 6-dimensional worldvolume Hodge star, and $v, w$ are orthogonal vectors used to describe the distribution of the dissolved D3 charge. 

In the description of D3-NS5 branes above, we have not restricted the range of $\theta \in (0, 2 \pi)$. For the construction of KPV state, we are interested in anti-D3-NS5 branes, which corresponds to the range $\theta \in (\pi, 3 \pi/4)$ of our description\footnote{The statement that anti-D3-NS5 branes are described by $\theta$ in the regime of  $(\pi, 3 \pi/4)$ is only strictly true for background where Maxwell charges and Page charges are the same.}. For convenience, we can do a reparametrisation $\theta \rightarrow \theta - \pi$ to bring it to the regime $\theta \in (0, \pi/2)$. In the new $\theta$, our currents are given by
\begin{align}
J_2 &= C r_h^2 \sin \theta \cos \theta \ v \wedge w \\
J_4 &=   C r_h^2 \sin \theta \ * (  v \wedge w) \\
j_6 &= - C r_h^2 \cos \theta \ * ( 1) 
\end{align}
where we have drop the superscript $equiv$ for syntactical simplicity.

\section{Blackfold perturbation equations} \label{secB}

In this appendix, we shall derive the blackfold perturbation equations for deformations around the KPV state. We start with a discussion of embedding geometry and computations of some useful variational expressions. Subsequently, we present the derivation of the blackfold perturbation equations used in the main text. For further discussion on embedding geometry and blackfold perturbation equation, see \cite{Carter:2000wv,Armas:2017pvj,Armas:2019iqs}.

\subsection{Useful definitions \& formulae}
\paragraph{Definitions} Given a manifold $\mathcal{M}$ and a submanifold $\mathcal{W}$ defined by the embedding $X^{\mu} (\sigma^a)$, we can define the induced metric
\be
\label{27}
\gamma_{ab} \equiv \p_a X^\mu \p_b X^\nu g_{\mu \nu}
\ee
the tangential projector
\be
h^{\mu \nu} \equiv \gamma^{ab} \p_a X^\mu \p_b X^\nu
\ee
and the orthogonal projector
\be
\perp_{\mu \nu} \equiv g_{\mu \nu} - h_{\mu \nu}
\ee
For convenience, let us define the object $\p_a X^\mu$ as
\be
\p^a X_\mu \equiv g_{\mu \nu} \gamma^{a b} \p_b X^\nu   
\ee
then the pullback of a general tensor from $\mathcal{M}$ to $\mathcal{W}$ is given by
\be
T^{a_1 a_2 ... a_n}_{\,\,\, \,\,\, \,\,\, \,\,\, \,\,\, \,\,\, \,\,\, \,\,\, b_1 b_2 ... b_m} \equiv \p^{a_1} X_{\mu_1} ... \ \p_{b_1} X^{\nu_1} ... \ T^{\mu_1 ... \mu_n}_{\,\,\, \,\,\, \,\,\, \,\,\, \,\,\, \,\, \nu_1 ... \nu_m}
\ee
Let us define also the extrinsic curvature  
\be
K_{\mu \nu}^{\,\,\, \,\,\ \rho} \equiv h^\sigma_\nu \overline{\nabla}_\mu  h^\rho_\sigma = - h^\sigma_\nu \overline{\nabla}_\mu \perp^\rho_\sigma
\ee
where $\overline{\nabla}_\mu = h^\rho_\mu \nabla_\rho$. By substitutions, we can show that
\be
\label{28}
K_{ab}^{\,\,\, \,\,\,\rho} = \p_a X^\mu \p_b X^\nu K_{\mu \nu}^{\,\,\, \,\,\ \rho} =  \nabla_a \left( \p_b X^\rho \right) + \Gamma^\rho_{\mu \nu} \p_a X^\mu \p_b X^\nu
\ee
where $\nabla_a$ acts only on the $b$ index of $\p_b X^\rho$: $\nabla_a (\p_b X^\rho) = \p_a (\p_b X^\rho) - \Theta_{a b}^c \p_c X^\rho$ with $\Theta_{a b}^c$ the Christoffel symbols of the induced metric $\gamma_{ab}$.

\paragraph{Variation of induced metric} Hitting $\delta$ to the definition of $\gamma_{ab}$ in (\ref{27}), we obtain the expression
\be
\delta \gamma_{a b} = \p_a X^\mu \p_b X^\nu \Big( \nabla_\mu \left( \delta X^\alpha g_{\alpha \nu} \right) + \nabla_\nu \left( \delta X^\alpha g_{\alpha \mu} \right) \Big)
\ee
When we embed a surface without edges in a higher dimensional background, the variations along the brane directions of the embedding functions $X^\mu (\sigma)$ can be cancelled by a reparametrisation of the worldvolume coordinates. As a result, we only have to worry about the variations of the transverse scalars $\delta X^\mu_{\perp} (\sigma)$ (i.e. $\p^a X_\mu \delta X^\mu_\perp = 0$). Making use of equation (\ref{28}), we have
\be
\delta \gamma_{ab} = - 2 K_{ab}^{\,\,\, \,\,\, \rho} \left( \delta X^\alpha_{\perp} g_{\alpha \rho} \right)
\ee
Using the identity $\gamma_{ab} \gamma^{bc} = \delta_a^c$, we can easily deduce that
\be
\delta \gamma^{ab} = 2 K^{ab}_{\,\,\, \,\,\, \rho} \delta X^\rho_{\perp} 
\ee

\paragraph{Variation of normal vectors} We note that the normal vectors are implicitly defined by
\begin{align}
\p_a X^\rho n_\rho^{(i)} &= 0\\
n_\rho^{(i)} n^\rho_{(j)} &= \delta^{(i)}_{\ (j)}
\end{align}
Hitting $\delta$ to both equations yields respectively the variation of $n^{(i)}_\rho$ along the worldvolume directions and normal to the worldvolume directions\footnote{As normal vectors are used collectively to specify the position of the branes inside the background, it's obvious that we have a rotational gauge symmetry in defining these vectors. Therefore, we can safely ignore variations regarding rotations of the normal vectors among themselves.}. 
\begin{align}
h^\rho_\sigma \, \delta n_\rho^{(i)} &= - \p^a X_\sigma \p_a \delta X^\rho_\perp n_\rho^{(i)} \\
\perp^{\rho}_\sigma \delta n_{\rho}^{(i)} &=  \frac{1}{2} n^{\alpha \, (i)} n^{\beta \, (i)} \p_\gamma g_{\alpha \beta} \delta X^\gamma_\perp n_\sigma^{(i)}
\end{align}
All together, we have 
\be
\label{30}
\delta n^{(i)}_\rho = - \p^a X_\rho \p_a \delta X^\sigma_\perp n_\sigma^{(i)} + \frac{1}{2} n^{\alpha \, (i)} n^{\beta \, (i)} \p_\gamma g_{\alpha \beta} \delta X^\gamma_\perp n_\rho^{(i)}
\ee

\paragraph{Variation of extrinsic curvature} Hitting $\delta$ to the expression of $K_{ab}^{\ \ \rho}$ in (\ref{28}), we obtain
\begin{equation}
\label{31}
\delta K_{ab}^{\,\,\, \,\,\, \rho} = \nabla_a \left( \p_b \delta X^\rho_\perp  \right) - \delta \Theta^c_{ab} \p_c X^\rho + \delta \Gamma^{\rho}_{\mu \nu} \p_a X^\mu \p_b X^\nu + 2 \Gamma^\rho_{\mu \nu} \p_a \delta X^\mu_\perp \p_b X^\nu
\end{equation}
Considering the variation of the projected extrinsic curvature $K_{ab}^{\ \ \, (i)}$, we have
\be
\delta \left( K_{ab}^{\,\,\, \,\,\, (i)} \right) = \delta \left( K_{ab}^{\,\,\, \,\,\, \rho} n_\rho^{(i)} \right) = \delta \left( K_{ab}^{\,\,\,\,\,\, \rho}  \right) n_{\rho}^{(i)} + K_{ab}^{\,\,\,\,\,\, \rho} \delta \left( n_\rho^{(i)} \right)
\ee
Making use of results in (\ref{30}) and (\ref{31}), we can write 
\begin{multline}
\label{42}
\delta\left( K_{ab}^{\,\,\, \,\,\, (i)}  \right) = n_\rho^{(i)} \nabla_a \left( \p_b \delta X^\rho_\perp \right) + n^{(i)}_\rho \delta X^\alpha_\perp \p_\alpha \Gamma^\rho_{\mu \nu} \p_a X^\mu \p_b X^\nu + 2 \, n_\rho^{(i)}  \Gamma^\rho_{\mu \nu} \p_a \delta X^\mu_\perp \p_b X^\nu \\
+ \frac{1}{2} K_{ab}^{\,\,\,\,\,\, \rho} \left( n^{\alpha \, (i)} n^{\beta \, (i)} \p_\gamma g_{\alpha \beta} \delta X^\gamma_\perp n_\rho^{(i)} \right)
\end{multline}

\paragraph{Variation of anti-D3-NS5 blackfold energy-momentum tensor} Hitting $\delta$ to the expression of $T^{ab}$ in (\ref{33}), we obtain the expression
\begin{multline}
\label{43}
\delta T^{a b} = - \mathbb{Q}_5 \sin \theta \delta ( \tan \theta )  \gamma^{ab} - \mathbb{Q}_5 \frac{1}{\cos\theta} \left( 2 K^{ab}_{\ \ \, \rho} \delta X^\rho_\perp \right) \\
+ \mathbb{Q}_5  \Big( \delta (v^a) v^b + v^a \delta (v^b) + \delta(w^a) w^b + w^a \delta (w^b) \Big) \tan \theta \sin \theta \\
+ \mathbb{Q}_5 (v^a v^b + w^a w^b) \sin \theta \delta  ( \tan \theta ) +  \mathbb{Q}_5 (v^a v^b + w^a w^b) \sin \theta \cos^2 \theta \delta (\tan \theta)
\end{multline}
We can also provide the general expressions for the variations of the blackfold currents. However, as the blackfold currents either enter our equations with a Hodge dual or coupled to the background fluxes, let us write down only the needed components when we use them.

\subsection{Current conservation equations}
Recall from (\ref{34})-(\ref{35}) the blackfold current conservation equations
\begin{align}
d * j_6 &= 0 \\
d * J_4 - * j_6 \wedge F_3 &= 0 \\
d * J_2 + H_3 \wedge * J_4 &= 0 
\end{align}
\begin{enumerate}
\item Considering the $j_6$ conservation equation, we can easily show that it gives rise to the perturbation equation
\be 
\label{A}
\p_a \delta \mathbb{Q}_5 = 0
\ee
where we have used $ * j_6 = \mathbb{Q}_5$.

\item Considering the $J_4$ conservation equation, firstly, we note that it can be rewritten as
\be
d * \tilde{J}_4 = 0
\ee 
where
\begin{align}
* \tilde{J}_4 &= * J_4 - *j_6 \wedge C_2\\
&=  - C r_h^2 \sin \theta \, v \wedge w - C r_h^2 \cos \theta \, C_2
\end{align}
From the unitary condition $v^a v_a = w^a w_a = 1$, it can be easily shown that
\begin{align}
&\delta v_\omega = \sqrt{M} b_0 \cos \psi & &\delta w_\varphi =  \sqrt{M} b_0 \cos \psi \sin \omega
\end{align}
Therefore, we have
\begin{multline}
\delta \left( * \tilde{J}_4 \right) = - \mathbb{Q}_5 \delta \tan \theta \, v \wedge w - \mathbb{Q}_5 \tan \theta \left( \delta v \wedge w + v \wedge \delta w \right) - \mathbb{Q}_5 \, \delta C_2 \\
= -\Bigg( \mathbb{Q}_5 M b_0^2 \sin^2 \psi \delta \tan \theta + 2 \mathbb{Q}_5 M b_0^2 \tan \theta \cos \psi \sin \psi \delta \psi + 2  \mathbb{Q}_5 M \sin^2 \psi \delta \psi \Bigg)  \sin \omega d \omega \wedge d \varphi \\
- \Big(\mathbb{Q}_5 \tan \theta \sqrt{M} b_0 \sin \psi \delta w_t\Big) d \omega \wedge dt  - \Big( \mathbb{Q}_5 \tan \theta \sqrt{M} b_0 \sin \psi  \sin \omega \delta v_t \Big) dt \wedge d \varphi
\end{multline}
where we have used that $C_2$ at the tip is given by $C_2 = M (\psi - \frac{1}{2} \sin 2 \psi) \sin \omega d \omega \wedge d \varphi$ and corrections away from the tip start at order $\mathcal{O} \left( r^2 \right)$. Thus, the $J_4$ perturbation equation is given by
\begin{multline}
\label{B}
- \mathbb{Q}_5 M b_0^2 \sin^2 \psi \sin \omega
\Bigg(  \p_t \delta \tan \theta + 2 \tan \theta \cot \psi \p_t \delta \psi + \frac{2}{b_0^2}  \p_t \delta \psi \Bigg) \\
=  \mathbb{Q}_5  M^{3/2} b^3_0 \tan \theta  \sin \psi \Big( \p_\varphi \delta w^t + \p_\omega  \left( \sin \omega \delta v^t \right) \Big)
\end{multline}
where we have used $\delta v_t = - M b_0^2 \, \delta v^t$ and $\delta w_t = - M b_0^2 \, \delta w^t$.

\item Considering the $J_2$ conservation equation, we have the variation of $* J_2$ is given by 
\begin{multline}
\delta \big( * J_2 \big) =  \mathbb{Q}_5 \left( \delta \sin \theta \right) * (v \wedge w)+ \mathbb{Q}_5 \sin \theta \delta \left( * (v \wedge w) \right) \\
= \mathbb{Q}_5 \left( \cos^3 \theta \delta \tan \theta \right) \sqrt{-\gamma} \Big(v^\omega w^\varphi d t \wedge ... \wedge dx_3  \Big) - 2 \mathbb{Q}_5 \sin \theta ( \sqrt{-\gamma} \gamma_{\omega \omega} K^{\omega \omega}_{\ \ \, \psi} \delta \psi) \Big(v^\omega w^\varphi d t \wedge ... \wedge dx_3  \Big)\\
+ \mathbb{Q}_5 \sin \theta \sqrt{-\gamma} \Big( \delta v^t w^\varphi d x_1 \wedge ... \wedge dx_3 \wedge d \omega + v^\omega \delta w^t d x_1 \wedge ... d x_3 \wedge d \varphi\\
+ \delta v^\omega w^\varphi d t \wedge ... \wedge dx_3 + v^\omega \delta w^\varphi d t \wedge ... \wedge dx_3   \Big)
\end{multline}

As $\delta \left( H_3 \wedge * J_4 \right) = \delta H_3 \wedge * J_4 + H_3 \wedge \delta (* J_4) = 0$, the $J_2$ perturbation equation is equivalent to the set of equations
\begin{align}
\label{C}
\cot \theta \cos^2 \theta \p_\omega \delta \tan \theta + \sqrt{M} b_0 \sin \psi \p_t \delta v^t  = 0\\
\cot \theta \cos^2 \theta  \p_\varphi \delta \tan \theta +  \sqrt{M} b_0 \sin \psi \sin \omega \p_t \delta w^t = 0 \\
\p_\varphi \delta v^t - \p_\omega (\sin \omega  \delta w^t )= 0
\end{align}
where we have used (\ref{5})-(\ref{6}).

\end{enumerate}

\subsection{Energy-momentum conservation equations}\label{energy-momentum}
Recall from (\ref{3000})-(\ref{3001}), the intrinsic and extrinsic blackfold equations 
\begin{align}
\nabla_a T^{a b} &= \p^b X_\mu \, \mathcal{F}^\mu \\
T^{ab} K_{ab}^{\,\,\, \,\,\, (i)}  &= \mathcal{F}^\mu \, n^{(i)}_\mu
\end{align}
where $\mathcal{F}^\mu$ denotes the force terms coming from the coupling of the currents to the fluxes (\ref{90}).

\subsubsection{Intrinsic perturbation equation} \label{intrinsic}
The blackfold intrinsic perturbation equation is given by
\be
\delta \left( \nabla_a T^{a b} \right) = \delta \left( \p^b X_\mu \, \mathcal{F}^\mu \right)
\ee
Considering the LHS, we have
\be
\delta \left( \nabla_a T^{ab} \right) = \nabla_a \delta T^{ab} - T^{b c} \nabla_c \left( K_\rho \delta X^\rho_\perp \right) - 2 T^{ac} \nabla_c \left( K_{a \ \rho}^{\ b} \delta X^\rho_\perp \right) + T^{ac} \nabla^b \left( K_{ac \rho} \delta X^\rho_\perp \right)
\ee
where $K^\rho = \gamma^{ab} K_{ab}^{\ \ \rho}$ and we have used the identity
\be
\delta \, \Theta^{b}_{ac} = \frac{1}{2} \gamma^{b d} (\nabla_a  \delta \gamma_{c d} + \nabla_c  \delta \gamma_{a d} - \nabla_d  \delta \gamma_{ac}) 
\ee
Considering the RHS, we have
\begin{align}
\delta \left( \p^b X_\mu \mathcal{F}^\mu \right) &= \delta \left(  \p^b X_\mu \right) \mathcal{F}^\mu +  \p^b X_\mu \delta \left(\mathcal{F}^\mu \right) \\
&=  \gamma^{t b} g_{\psi \psi} \p_t \delta \psi  \left( F_3^{\psi \omega \varphi} J_{2 \omega \varphi} \right)
\end{align}
where we have made use of the explicit expression of $\mathcal{F}^\mu$ in (\ref{90}). Altogether, we have the intrinsic perturbation equation
\begin{multline}
\nabla_a \delta T^{ab} - T^{b c} \nabla_c \left( K_\rho \delta X^\rho_\perp \right) - 2 T^{ac} \nabla_c \left( K_{a \ \rho}^{\ b} \delta X^\rho_\perp \right) + T^{ac} \nabla^b \left( K_{ac \rho} \delta X^\rho_\perp \right)\\
=  \gamma^{t b} g_{\psi \psi} \p_t \delta \psi \left( F_3^{\psi \omega \varphi} J_{2 \omega \varphi} \right)
\end{multline}
Substituting in appropriate expressions, we obtain for $b = t,\omega, \varphi$ respectively 

\begin{enumerate}
\item The $t$ intrinsic perturbation equation  
\begin{multline}
\label{D}
\p_t \delta \tan \theta +  \frac{\sqrt{M} b_0}{\sin \psi} \tan \theta \left( \p_\omega \delta v^t + \frac{1}{\sin \omega} \p_\varphi \delta w^t + \cot \omega \delta v^t \right) 
 \\
 + 2 \left( \cot \psi \tan \theta + \frac{1}{b_0^2} \right) \p_t \delta \psi = 0
\end{multline} 

\item The $\omega$ intrinsic perturbation equation 
\be
\sqrt{M} b_0 \sin \psi  \tan^2 \theta  \p_t \delta v^t +   \sin \theta \cos \theta  \p_\omega \delta \tan \theta    = 0
\ee

\item The $\varphi$ intrinsic perturbation equation 
\be
\sqrt{M} b_0 \sin \psi \sin \omega \tan^2 \theta \p_t \delta w^t  + \sin \theta \cos \theta \p_\varphi \delta \tan \theta  = 0
\ee
\end{enumerate}

\subsubsection{Extrinsic equation} \label{extrinsic}
The extrinsic blackfold perturbation equation is given by
\be
\delta \left( T^{ab} K_{ab}^{\,\,\, \,\,\, (i)} \right) = \delta \left( \mathcal{F}^\mu \, n^{(i)}_\mu \right)
\ee
Making use of the results in (\ref{42}), we can easily write the LHS as
\begin{multline}
\delta \left( T^{ab} K_{ab}^{\,\,\, \,\,\, (i)} \right) = \delta T^{ab} K_{ab}^{\ \ \, (i)} + T^{ab} n_\rho^{(i)} \nabla_a \left( \p_b \delta X^\rho_\perp \right) + T^{ab} n^{(i)}_\rho \delta X^\alpha_\perp \p_\alpha \Gamma^\rho_{\mu \nu} \p_a X^\mu \p_b X^\nu\\
+ 2 \,T^{ab} n_\rho^{(i)}  \Gamma^\rho_{\mu \nu} \p_a \delta X^\mu_\perp \p_b X^\nu + \frac{1}{2} T^{ab} K_{ab}^{\,\,\,\,\,\, \rho} \left( n^{\alpha \, (i)} n^{\beta \, (i)} \p_\gamma g_{\alpha \beta} \delta X^\gamma_\perp n_\rho^{(i)} \right)
\end{multline}
For our purpose, we are interested in the orthogonal directions $\psi$ and $r$. The unitary normal vectors specifying these directions are respectively
\begin{align}
&n^{(1)} = \sqrt{M} b_0 d \psi & &n^{(2)} = \sqrt{M} b_0 d r 
\end{align}
For the $\psi$ direction, the RHS is given by 
\be
\delta \left( \mathcal{F}^\mu n_\mu^{(1)} \right) = \delta  \mathcal{F}^\mu n_\mu^{(1)} + \mathcal{F}^\mu \delta n_\mu^{(1)} = \delta \mathcal{F}^\psi n_\psi^{(1)}
\ee
The expression of $\delta \mathcal{F}^\psi$ can be easily obtained by hitting $\delta$ to the force term $\mathcal{F}^\mu$ (\ref{90}). As the computation is tedious but straight-forward, we shall not include all the details here. Nevertheless, for the convenience of the readers, let us note down the final results along with some useful (non-vanishing) intermediate steps. We have
\begin{align}
\delta F_3^{\psi \omega \varphi} &= \delta \left( g^{\psi \mu} \gamma^{\omega a_1} \p_{a_1} X^{\alpha_1} \gamma^{\varphi a_2} \p_{a_2} X^{\alpha_2} F_{3 \mu \alpha_1 \alpha_2} \right) \\
&= g^{\psi \psi} \left( \delta  \gamma^{\omega \omega} \right) \gamma^{\varphi \varphi} F_{3 \psi \omega \varphi} + g^{\psi \psi} \gamma^{\omega \omega} \left( \delta  \gamma^{\varphi \varphi} \right) F_{3 \psi \omega \varphi} + g^{\psi \psi} \gamma^{\omega \omega} \gamma_{\varphi \varphi} \left( \delta  F_{3 \psi \omega \varphi}  \right) \\
&= \left( 4 g^{\psi \psi} K^{\omega \omega}_{\ \ \ \psi} \gamma^{\varphi \varphi} F_{3 \psi \omega \varphi} + g^{\psi \psi} \gamma^{\omega \omega} \gamma^{\varphi \varphi}  \p_\psi F_{3 \psi \omega \varphi} \right) \delta \psi
\end{align}
Similarly, we have 
\be
\delta H_{7}^{\psi t ... \varphi} = \left( 4 g^{\psi \psi} \gamma^{tt} ... \gamma^{x^3 x^3} K^{\omega \omega}_{\ \ \ \psi} \gamma^{\varphi \varphi} H_{7 \psi t ... \varphi} + g^{\psi \psi} \gamma^{tt}... \gamma^{\varphi \varphi} \p_\psi H_{7 \psi t ... \varphi} \right) \delta \psi
\ee
Let us note also that 
\begin{align}
\delta J_{2 \omega \varphi} &= \mathbb{Q}_5 \left( \delta \sin \theta \right) v_\omega w_\varphi + \mathbb{Q}_5 \sin \theta \left( \delta v_\omega w_\varphi + v_\omega \delta w_\varphi \right) \\
&= \left( M b_0^2 \mathbb{Q}_5 \cos^3 \theta  \sin^2 \psi \sin \omega \right) \delta \tan \theta + \left( 2  M b_0^2 \mathbb{Q}_5 \sin \theta \cos \psi \sin \psi \sin \omega \right) \delta \psi
\end{align}
and
\begin{align}
\delta j_{6 t ... \varphi} &= - \mathbb{Q}_5 \left( \delta \sqrt{-\gamma} \right) = - \frac{1}{2} \mathbb{Q}_5 \sqrt{-\gamma} \gamma^{\alpha \beta} \delta \gamma_{\alpha \beta} \\
&=  \left( 2 \, \mathbb{Q}_5 \sqrt{-\gamma} \gamma^{\omega \omega} K_{\omega \omega}^{\ \ \ \psi} g_{\psi \psi} \right) \delta \psi 
\end{align}
Altogether, we have the variation of the force term $\delta \mathcal{F}^\psi$ is given by
\be
\delta \mathcal{F}^\psi = - \left( \delta H_7^{\psi t ... \varphi} \right) j_{6 t ... \varphi} - H_7^{\psi t ... \varphi} \left( \delta j_{6 t ... \varphi} \right) + \left( \delta F_3^{\psi \omega \varphi} \right) J_{2 \omega \varphi} + F_3^{\psi \omega \varphi} \left( \delta J_{2 \omega \varphi} \right)
\ee
For the $r$ direction, the RHS is given by
\be
\delta \left( \mathcal{F}^\mu n_\mu^{(2)} \right) = \delta  \mathcal{F}^\mu n_\mu^{(2)} + \mathcal{F}^\mu \delta n_\mu^{(2)} = \delta \mathcal{F}^r n_r^{(2)}
\ee
Similar to our treatment of $\delta \mathcal{F}^\psi$, we shall not present here the full computation of $\delta \mathcal{F}^r$ but only the final results along with some useful (non-vanishing) intermediate steps. We have
\begin{align}
\delta \tilde{F}_5^{r t ... x^3} &= \delta \left( g^{r \nu} \gamma^{t a_1} ... \gamma^{x^3 a_4} \p_{a_1} X^{\alpha_1} ... \p_{a_4} X^{\alpha_4} \tilde{F}_{5 \nu \alpha_1 ... \alpha_4} \right) \\
&= \left(  g^{rr} \gamma^{tt} ... \gamma^{x^3 x^3}  \p_r \tilde{F}_{5 r t ... x^3} \right) \delta r
\end{align}
The variation of the force term $\delta \mathcal{F}^r$ is given by
\be
\delta \mathcal{F}^r = \left( \delta \tilde{F}_5^{r t ... x^3} \right) J_{4 t ... x^3}
\ee
Substituting in appropriate expressions and simplify where possible, we obtain respectively 
\begin{enumerate}
\item The $\psi$ extrinsic perturbation equation
\be
\label{E}
(\p_t)^2 \delta \psi -  \frac{ \cos^2 \theta}{ \sin^2 \psi}   \nabla^2 \delta \psi  = \frac{2  \cos^2 \theta}{ \sin^2 \psi}  \delta \psi + \frac{2}{b_0^2}  \cos^2 \theta \left( 1 + \sin \theta \right) \delta \tan \theta
\ee

\item The $r$ extrinsic perturbation equation 
\begin{multline}
 (\p_t)^2 \delta r - \frac{\cos^2 \theta}{ \sin^2 \psi}  \nabla^2 \delta r   = \frac{8 a_2}{a_0} \sin \theta \delta r +  \frac{8 a_2 }{ a_0}  \delta r   -  \frac{  16 a_0 + 20 a_2  }{5  a_0} \cos^2 \theta  \delta r \\
 +  \frac{ 4  }{5} \cos^2 \theta   \sin^2 \omega  \delta r 
\end{multline}
where $\nabla^2$ is the normalised Laplacian, i.e. $\nabla^2 = (\p_\omega)^2 + 1/\sin^2 \omega (\p_\varphi)^2 + \cot \omega \p_\omega$. 
\end{enumerate}

\end{appendices}

\bibliographystyle{JHEP}
\bibliography{ref}

\providecommand{\href}[2]{#2}\begingroup\raggedright\begin{thebibliography}{10}

\bibitem{Kachru2003DeTheory}
S.~Kachru, R.~Kallosh, A.~D. Linde, and S.~P. Trivedi, {\it {De Sitter vacua in
  string theory}},  {\em Phys. Rev.} {\bf D68} (2003) 046005,
  [\href{http://arxiv.org/abs/hep-th/0301240}{{\tt hep-th/0301240}}].

\bibitem{Danielsson:2018ztv}
U.~H. Danielsson and T.~Van~Riet, {\it {What if string theory has no de Sitter
  vacua?}},  {\em Int. J. Mod. Phys.} {\bf D27} (2018), no.~12 1830007,
  [\href{http://arxiv.org/abs/1804.01120}{{\tt arXiv:1804.01120}}].

\bibitem{Kachru2002Brane/fluxTheory}
S.~Kachru, J.~Pearson, and H.~L. Verlinde, {\it {Brane / flux annihilation and
  the string dual of a nonsupersymmetric field theory}},  {\em JHEP} {\bf 06}
  (2002) 021, [\href{http://arxiv.org/abs/hep-th/0112197}{{\tt
  hep-th/0112197}}].

\bibitem{Klebanov2000SupergravitySingularities}
I.~R. Klebanov and M.~J. Strassler, {\it {Supergravity and a confining gauge
  theory: Duality cascades and chi SB resolution of naked singularities}},
  {\em JHEP} {\bf 08} (2000) 052,
  [\href{http://arxiv.org/abs/hep-th/0007191}{{\tt hep-th/0007191}}].

\bibitem{Bena2010OnKlebanov-Strassler}
I.~Bena, M.~Grana, and N.~Halmagyi, {\it {On the Existence of Meta-stable Vacua
  in Klebanov-Strassler}},  {\em JHEP} {\bf 09} (2010) 087,
  [\href{http://arxiv.org/abs/0912.3519}{{\tt arXiv:0912.3519}}].

\bibitem{Bena:2014jaa}
I.~Bena, M.~Graña, S.~Kuperstein, and S.~Massai, {\it {Giant Tachyons in the
  Landscape}},  {\em JHEP} {\bf 02} (2015) 146,
  [\href{http://arxiv.org/abs/1410.7776}{{\tt arXiv:1410.7776}}].

\bibitem{Michel:2014lva}
B.~Michel, E.~Mintun, J.~Polchinski, A.~Puhm, and P.~Saad, {\it {Remarks on
  brane and antibrane dynamics}},  {\em JHEP} {\bf 09} (2015) 021,
  [\href{http://arxiv.org/abs/1412.5702}{{\tt arXiv:1412.5702}}].

\bibitem{Bena:2016fqp}
I.~Bena, J.~Blåbäck, and D.~Turton, {\it {Loop corrections to the antibrane
  potential}},  {\em JHEP} {\bf 07} (2016) 132,
  [\href{http://arxiv.org/abs/1602.05959}{{\tt arXiv:1602.05959}}].

\bibitem{Cohen-Maldonado:2015ssa}
D.~Cohen-Maldonado, J.~Diaz, T.~van Riet, and B.~Vercnocke, {\it {Observations
  on fluxes near anti-branes}},  {\em JHEP} {\bf 01} (2016) 126,
  [\href{http://arxiv.org/abs/1507.01022}{{\tt arXiv:1507.01022}}].

\bibitem{Armas:2018rsy}
J.~Armas, N.~Nguyen, V.~Niarchos, N.~A. Obers, and T.~Van~Riet, {\it
  {Metastable Nonextremal Antibranes}},  {\em Phys. Rev. Lett.} {\bf 122}
  (2019), no.~18 181601, [\href{http://arxiv.org/abs/1812.01067}{{\tt
  arXiv:1812.01067}}].

\bibitem{Niarchos:2015moa}
V.~Niarchos, {\it {Open/closed string duality and relativistic fluids}},  {\em
  Phys. Rev.} {\bf D94} (2016), no.~2 026009,
  [\href{http://arxiv.org/abs/1510.03438}{{\tt arXiv:1510.03438}}].

\bibitem{Bena:2015kia}
I.~Bena and S.~Kuperstein, {\it {Brane polarization is no cure for tachyons}},
  {\em JHEP} {\bf 09} (2015) 112, [\href{http://arxiv.org/abs/1504.00656}{{\tt
  arXiv:1504.00656}}].

\bibitem{Emparan:2009cs}
R.~Emparan, T.~Harmark, V.~Niarchos, and N.~A. Obers, {\it {World-Volume
  Effective Theory for Higher-Dimensional Black Holes}},  {\em Phys. Rev.
  Lett.} {\bf 102} (2009) 191301, [\href{http://arxiv.org/abs/0902.0427}{{\tt
  arXiv:0902.0427}}].

\bibitem{Emparan:2009at}
R.~Emparan, T.~Harmark, V.~Niarchos, and N.~A. Obers, {\it {Essentials of
  Blackfold Dynamics}},  {\em JHEP} {\bf 03} (2010) 063,
  [\href{http://arxiv.org/abs/0910.1601}{{\tt arXiv:0910.1601}}].

\bibitem{Armas:2016mes}
J.~Armas, J.~Gath, V.~Niarchos, N.~A. Obers, and A.~V. Pedersen, {\it {Forced
  Fluid Dynamics from Blackfolds in General Supergravity Backgrounds}},  {\em
  JHEP} {\bf 10} (2016) 154, [\href{http://arxiv.org/abs/1606.09644}{{\tt
  arXiv:1606.09644}}].

\bibitem{M2M5brane}
J.~Armas, N.~Nguyen, V.~Niarchos, and N.~A. Obers, {\it {Thermal transitions of
  metastable M-branes}},  {\em JHEP} {\bf 08} (2019) 128,
  [\href{http://arxiv.org/abs/1904.13283}{{\tt arXiv:1904.13283}}].

\bibitem{Bhattacharyya:2008jc}
S.~Bhattacharyya, V.~E. Hubeny, S.~Minwalla, and M.~Rangamani, {\it {Nonlinear
  Fluid Dynamics from Gravity}},  {\em JHEP} {\bf 02} (2008) 045,
  [\href{http://arxiv.org/abs/0712.2456}{{\tt arXiv:0712.2456}}].

\bibitem{Camps:2012hw}
J.~Camps and R.~Emparan, {\it {Derivation of the blackfold effective theory}},
  {\em JHEP} {\bf 03} (2012) 038, [\href{http://arxiv.org/abs/1201.3506}{{\tt
  arXiv:1201.3506}}]. [Erratum: JHEP06,155(2012)].

\bibitem{Klebanov:2010qs}
I.~R. Klebanov and S.~S. Pufu, {\it {M-Branes and Metastable States}},  {\em
  JHEP} {\bf 08} (2011) 035, [\href{http://arxiv.org/abs/1006.3587}{{\tt
  arXiv:1006.3587}}].

\bibitem{Cvetic:2000db}
M.~Cvetic, G.~Gibbons, H.~Lu, and C.~Pope, {\it {Ricci flat metrics, harmonic
  forms and brane resolutions}},  {\em Commun. Math. Phys.} {\bf 232} (2003)
  457--500, [\href{http://arxiv.org/abs/hep-th/0012011}{{\tt hep-th/0012011}}].

\bibitem{Klebanov:2000hb}
I.~R. Klebanov and M.~J. Strassler, {\it {Supergravity and a confining gauge
  theory: Duality cascades and chi SB resolution of naked singularities}},
  {\em JHEP} {\bf 08} (2000) 052,
  [\href{http://arxiv.org/abs/hep-th/0007191}{{\tt hep-th/0007191}}].

\bibitem{Herzog:2001xk}
C.~P. Herzog, I.~R. Klebanov, and P.~Ouyang, {\it {Remarks on the warped
  deformed conifold}},  in {\em {Modern Trends in String Theory: 2nd Lisbon
  School on g Theory Superstrings Lisbon, Portugal, July 13-17, 2001}}, 2001.
\newblock \href{http://arxiv.org/abs/hep-th/0108101}{{\tt hep-th/0108101}}.

\bibitem{Minasian:1999tt}
R.~Minasian and D.~Tsimpis, {\it {On the geometry of nontrivially embedded
  branes}},  {\em Nucl. Phys. B} {\bf 572} (2000) 499--513,
  [\href{http://arxiv.org/abs/hep-th/9911042}{{\tt hep-th/9911042}}].

\bibitem{Harmark:1999rb}
T.~Harmark and N.~A. Obers, {\it {Phase structure of noncommutative field
  theories and spinning brane bound states}},  {\em JHEP} {\bf 03} (2000) 024,
  [\href{http://arxiv.org/abs/hep-th/9911169}{{\tt hep-th/9911169}}].

\bibitem{Emparan:2011hg}
R.~Emparan, T.~Harmark, V.~Niarchos, and N.~A. Obers, {\it {Blackfolds in
  Supergravity and String Theory}},  {\em JHEP} {\bf 08} (2011) 154,
  [\href{http://arxiv.org/abs/1106.4428}{{\tt arXiv:1106.4428}}].

\bibitem{Marolf:2000cb}
D.~Marolf, {\it {Chern-Simons terms and the three notions of charge}},  in {\em
  {Quantization, gauge theory, and strings. Proceedings, International
  Conference dedicated to the memory of Professor Efim Fradkin, Moscow, Russia,
  June 5-10, 2000. Vol. 1+2}}, pp.~312--320, 2000.
\newblock \href{http://arxiv.org/abs/hep-th/0006117}{{\tt hep-th/0006117}}.

\bibitem{Carter:2000wv}
B.~Carter, {\it {Essentials of classical brane dynamics}},  {\em Int. J. Theor.
  Phys.} {\bf 40} (2001) 2099--2130,
  [\href{http://arxiv.org/abs/gr-qc/0012036}{{\tt gr-qc/0012036}}].

\bibitem{Armas:2017pvj}
J.~Armas and J.~Tarrio, {\it {On actions for (entangling) surfaces and DCFTs}},
   {\em JHEP} {\bf 04} (2018) 100, [\href{http://arxiv.org/abs/1709.06766}{{\tt
  arXiv:1709.06766}}].

\bibitem{Armas:2019iqs}
J.~Armas and E.~Parisini, {\it {Instabilities of Thin Black Rings: Closing the
  Gap}},  {\em JHEP} {\bf 04} (2019) 169,
  [\href{http://arxiv.org/abs/1901.09369}{{\tt arXiv:1901.09369}}].

\end{thebibliography}\endgroup


\providecommand{\href}[2]{#2}\begingroup\raggedright\endgroup

\end{document}